\documentclass[aps,prd,twocolumn,floatfix,noshowpacs,tightenlines,noshowkeys,superscriptaddress,amsmath,amssymb,
nofootinbib]{revtex4-1}
\usepackage{amssymb,amsbsy,epsfig,color,graphicx}
\usepackage{color}
\usepackage{[longtable}
\usepackage{array}
\usepackage{dcolumn}   
\usepackage{cellspace}
\usepackage{mathtools}
\usepackage{amstext}
\usepackage{amssymb}
\usepackage{stmaryrd}
\usepackage{stackrel}
\usepackage{graphicx}
\usepackage{esint}
\usepackage[utf8]{inputenc}
\usepackage{float}
\restylefloat{table}
\usepackage{booktabs}
\usepackage{enumitem}
\usepackage{slashed}
\usepackage{hyperref}
\usepackage{etoolbox} 
\usepackage{lipsum} 
\usepackage[capitalize]{cleveref}
\usepackage{multirow}
\usepackage[caption=false]{subfig}
\allowdisplaybreaks
\usepackage{soul}

\def\be{\begin{equation}}
\def\ee{\end{equation}}
\def\bea{\begin{eqnarray}}
\def\eea{\end{eqnarray}}

\newcommand{\ba}{\begin{eqnarray}}
\newcommand{\ea}{\end{eqnarray}}

\renewcommand\[{\begin{equation}}
\renewcommand\]{\end{equation}}
\makeatletter

\appto{\appendix}{%
\@ifstar{\def\theequation@prefix{A.}}%
{}%
}
\makeatother

\hypersetup{pdfstartview=FitV,colorlinks=true,linkcolor=midblue,citecolor=midblue,filecolor=midblue,urlcolor=midblue}

\definecolor{midblue}{rgb}{0,0,0.5}
\definecolor{cadmiumorange}{rgb}{0.93, 0.53, 0.18}

\def\RN{Reissner-Nordstr\"{o}m}

\def\Schld{Schwarzschild }
\def\dS{de Sitter }

\def\ro{\varrho}
\usepackage{hyperref}
\hypersetup{colorlinks=true}

\def\equationautorefname~#1\null{%
	Eq.~(#1)\null
}
\def\figureautorefname~#1\null{%
	Fig.~#1\null
}
\def\tableautorefname~#1\null{%
	Table.~#1\null
}
\def\sectionautorefname~#1\null{%
	Section #1\null
}
\def\appendixautorefname~#1\null{%
	Appendix #1\null
}
\begin{document}

\title{Ringdown of Charged Compact Objects using Membrane Paradigm}

\author{Mostafizur Rahman}
\email{mostafizur.r@iitgn.ac.in}
\affiliation{Indian Institute of Technology, Gandhinagar, Gujarat-382355, India}

\author{Arpan Bhattacharyya}
\email{abhattacharyya@iitgn.ac.in}
\affiliation{Indian Institute of Technology, Gandhinagar, Gujarat-382355, India}

\begin{abstract}
Although black holes are an integral part of the standard model of astrophysics and cosmology, their existence poses some serious fundamental problems. In recent years, several horizonless compact object models were proposed to address those issues. As the gravitational wave detectors started to observe more and more merger events with a large signal-to-noise ratio, gravitational wave spectroscopy could hold the key to uncover the existence of these objects. This is because the late time ringdown signals of horizonless compact objects differ from that of the black holes. In this paper, we study the ringdown properties of charged compact objects and compare them with those obtained in the black hole scenario. Since the internal structure and the equation of state of these compact objects are largely unknown, we employ membrane paradigm to obtain appropriate boundary conditions for the perturbations of these objects. This model can describe the ringdown properties of a large variety of compact objects.
\end{abstract}

\maketitle


\section{Introduction}\label{Intro}

The black holes are defined as the region of spacetime where the gravity is so strong that even light can not escape their crushing grip. They are widely accepted as dark astrophysical objects, which are the end stage of dying stars with a mass roughly greater than three solar masses \cite{Narayan:2005ie, Mueller:2007rz}. The idea of black holes is interesting for several reasons. The physics of gravitational collapse and the formation of black holes are well understood. Moreover, the spacetime outside the black hole is perfectly regular as curvature singularities are hidden inside its event horizon and thus do not hinder measurements outside the black hole region \cite{Wald:1984rg, Poisson:2009pwt}. Furthermore, it has been established that black holes in general relativity (GR) are stable against external perturbations \cite{PhysRev.108.1063,Klainerman:2017nrb,PhysRevD.9.860,Kay_1987,Dafermos:2008ys,Casals:2019vdb}. They are also the simplest astrophysical objects. ``No-hair'' theorem says that stationary black holes in GR can be characterized by their mass, charge and angular momentum \cite{https://doi.org/10.1002/asna.19752960110,PhysRev.164.1776,PhysRevLett.26.331}. These properties of black holes make them an ideal laboratory to probe physics in strong gravity regime \cite{Berti:2015itd,TheLIGOScientific:2016src,Yunes:2016jcc}. Recent discovery of gravitational waves (GWs) and black hole shadow shows progress in this direction \cite{Abbott:2016blz,LIGOScientific:2019fpa,Akiyama:2019cqa}.\\ 
 Nevertheless, the theory of black holes poses some serious questions to the conventional view on nature. Although the exterior of these objects is free from pathologies, the same can not be said about their interior. These objects shelter singularities and Cauchy horizon (for charged and rotating black holes ) in their interior where classical GR breaks down \cite{1978tpar.book..217P,Reall2018ViewpointAP,Dafermos:2003wr,Cardoso:2017soq,Dias:2018ynt,Rahman:2018oso}. Moreover, Hawking showed that these objects can emit radiations using semi-classical analysis \cite{1998bhp..book.....F}. However, complete evaporation of the black hole breaks the unitary property of quantum mechanics as it can turn a pure state into a mixed state\cite{Mathur:2009hf}. The resolution of these problems is essential to understand the quantum mechanical description of gravity.  Several horizonless dark exotic compact object (ECO) models were proposed to address some of those issues in recent years.  It has been conjectured that \cite{Mathur:2005zp,Bena:2007kg,Bena:2013dka,Skenderis:2008qn,Mathur:2008nj},  one can describe black holes as ensembles of micro-states represented by smooth, horizonless geometries without closed time-like curves, termed as fuzzballs. Sometimes, it is possible to find such horizonless solutions, with the same charges as a black hole from supergravity for eg. \cite{Rychkov:2005ji,Lunin:2001fv,Mathur:2003hj}.
 In the gravastar model, the event horizon is replaced by a thin shell that connects a \dS\ interior core with an external \Schld\ geometry \cite{Chapline:2000en,Mazur:2004fk,Visser:2003ge}. They can arise as a result of phase transition near what could have been the position of the event horizon. They can also originate through quantum field theoretical processes, which replace the horizon with a quantum phase boundary \cite{Mottola:2006ew,Mottola:2011ud,Kawai:2017txu}.
   Axion or axion like dark energy models predict the existance of dark compact objects, called the boson stars which can form naturally from gravitational collapse \cite{Brito:2015yga, Liebling:2012fv, Seidel:1993zk}. \par
 It is expected that the observational signatures of the ECOs will differ in general from those of the black holes. In order to study those signatures, we need to quantify the characteristics of these objects. Since the details about their internal structure are largely unknown, a model-independent way has been proposed by addressing the following questions: how compact and dark are these objects are \cite{Cardoso:2019rvt,Maggio:2020jml}?  These parameters allow us to pragmatically quantify the difference between ECO and black hole spacetime.
 From the observational perspective, gravitational wave spectroscopy could be a very powerful tool to distinguish these objects from black holes \cite{Cardoso:2016rao,Maggio:2020jml, Mark:2017dnq, Micchi:2019yze,Micchi:2020gqy}. Merger of compact object binaries are among the most important astrophysical processes that give rise to the GWs. It is widely believed that the post-merger remnant eventually settles down to a stationary state by emitting GWs through a process called ringdown. These waves not only contains information about the nature of the final object but also about the underlying theory of gravity \cite{Cardoso:2016oxy,Yunes:2013dva,Berti:2015itd,TheLIGOScientific:2016src,Yunes:2016jcc,Dey:2020lhq}. It has been noted in the literature that the ringdown properties of ECOs generally differ from that of the black holes. It is well-known that the ringdown signal of a black hole dies down at later times \cite{PhysRevD.34.384,PhysRevD.49.883,PhysRevD.49.890,PhysRevLett.84.10}. However, one obtains modulated repetitive signals from the compact object, so-called the \textit{gravitational wave echoes} \cite{Cardoso:2017cqb,Cardoso:2016rao,Cardoso:2019rvt,Maggio:2020jml, Mark:2017dnq, Micchi:2019yze,Micchi:2020gqy}. The repetition time is usually governed by the compactness parameter, whereas the modulation depends on the reflection coefficient of the object.  In fact, recently some authors claimed that they have found observational evidence of GW echoes in the LIGO data \cite{Conklin:2017lwb,PhysRevD.97.124044,Abedi:2016hgu,Holdom:2019bdv,Conklin:2019fcs}. However, these claims are still shrouded in controversy \cite{Tsang:2019zra, Salemi:2019uea, Uchikata:2019frs, Westerweck:2017hus}.\par
 One of the most important steps to study the ringdown properties of ECOs is to set up appropriate boundary conditions. In \cite{Oshita:2019sat, Wang:2019rcf} for objects like fuzzballs and gravastars, the authors used gauge/gravity duality and membrane paradigm to study it. In black hole membrane paradigm, the event horizon is viewed as $2+1$ dimensional viscous fluid membrane \cite{1982mgm..conf..587D, Thorne, MacDonald:1982zz, PhysRevD.33.915}. This formulation gives us an intuitive understanding of many of the properties of black holes. In \cite{Maggio:2020jml}, the authors used the membrane paradigm to obtain the boundary conditions for neutral ECOs in a model-independent way.  Ref.~\cite{Chen:2020htz} done the same for rotating compact objects.\par
 In this paper, we study the ringdown properties of charged compact objects using the membrane paradigm. Motivation for considering charged black holes stems from the fact that they provide us with a variety of interesting phenomenological scenarios, e.g., in the context of standard model \cite{Maldacena:2020skw}, in astrophysical context \cite{Mereghetti:2008je}. It is well known that electrically charged astrophysical objects get neutralized rather quickly by accreting ionized plasma and other quantum mechanical processes like Schwinger's pair production and Hawking radiation as a consequence of electron's large charge-to-mass ratio $e/m_e\approx10^{21}$ \cite{Cardoso:2016olt, Ghosh:2020tdu}. Nevertheless, certain minicharged dark matter models can evade the fast neutralization process since the charge of these minicharged particles is only a fraction of the electrons' charge \cite{Cardoso:2016olt}. Another interesting possibility is the magnetically charged black holes \cite{Maldacena:2020skw, Ghosh:2020tdu} and compact objects like magnetars \cite{Mereghetti:2008je}. It has been discussed in the literature that these objects are relatively long-lived than their electrically charged counterpart (see \cite{Ghosh:2020tdu} for further discussion). Since the construction discussed here is quite generic, it can be applied to a variety of compact object models \cite{Mathur:2005zp,Bena:2007kg,Bena:2013dka,Skenderis:2008qn,Mathur:2008nj,Chapline:2000en,Mazur:2004fk,Visser:2003ge,Brito:2015yga, Liebling:2012fv, Seidel:1993zk,Gimon:2007ur,Buoninfante:2019swn,Damour:2007ap,Roy:2019yrr} and could potentially shed some light to the observational signatures of some modified gravity theories e.g. the brane-world models \cite{Dadhich:2000am,Dey:2020lhq,Dey:2020pth,Chakraborty:2021gdf}. This construction is also interesting from the phenomenological point of view since it gives an intuitive understanding of the interactions of electromagnetic fields with the charged membrane.\par
The paper is organized as follows: In \autoref{Sec_Membrane}, we briefly discuss the membrane paradigm. In \autoref{Sec_Perturb}, we show how this paradigm leads to the appropriate boundary conditions of charged compacts objects. In \autoref{Sec_Ringdown} and \autoref{Sec_QNM}, we present the ringdown properties and QNMs of these objects. We discuss the quasinormal modes of the compact object in the limit $Q\to M$ in \autoref{Near extremal modes}. A discussion about the detectability of gravitation wave echoes is given \autoref{Detect}. Finally, with some remarks, we conclude our paper.\par
\textit{Notation and Convention}: Throughout the paper, positive signature convention $(-,+,+,+)$ is used. Greek indices are used to denote four-dimensional tensors, whereas lower case roman indices are used to denote three-dimensional tensors. The fundamental constants are assumed as $\hbar=c=G=1$.

 
\section{Membrane paradigm for perturbed charged compact objects}\label{Sec_Membrane}

The black hole membrane paradigm provides an intuitive way to understand the dynamics of a black hole. In the membrane paradigm, we consider a fictitious $2+1$ dimensional time-like surface just outside the black hole event horizon 
$\mathcal{H}$, so-called the stretched horizon $\mathcal{S}$ \cite{1982mgm..conf..587D, Thorne, MacDonald:1982zz, PhysRevD.33.915}. It can be thought of as the world tube of a family of fudicial observers (FIDO) which are at rest with respect to the black hole.   We denote the four velocities of the FIDOs by $u^{\mu}$. Thus, the time-like congruence $u^{\mu}$ is the generator of the stretched horizon. The outward-pointing unit normal to $\mathcal{S}$ is denoted by $n^{\mu}$. The energy-momentum tensor $\tau_{ab}$ of some matter sources at stretched horizon produces a jump in the extrinsic curvature $K^a_{~b}=h^c_{~b}n_{a;c}$ in accordance to Israel junction condition \cite{1966NCimB..44....1I,Darmois1927, PhysRevD.58.064011},
\begin{equation}\label{Israel_condi}
	\tau_{ab}=\big(\big[K\big]h_{ab}-\big[K_{ab}\big]\big)
\end{equation}
where, $K$ is the trace of the extrinsic curvature $K_{ab}$ and  $\big[K_{ab}\big]=K^{+}_{ab}-K^{-}_{ab}$ denotes its discontinuity across the stretched horizon. The metric tensor on $\mathcal{S}$ is represented by $h_{ab}=g_{ab}-n_{a}n_{b}$. In the membrane paradigm, the energy-momentum tensor is chosen to contain all the information about the interior \cite{PhysRevD.58.064011}. Hence the following condition is satisfied \cite{PhysRevD.58.064011, Maggio:2020jml}, 
\begin{eqnarray}\label{Membrane_paradigm}
&{K}^{-}_{ab}=0\,,\qquad{\tau}_{ab}=\big(K h_{ab}-K_{ab}\big)~
\end{eqnarray}
where, we have omitted the prefix in $K^{+}_{ab}$ for brevity. When the stretched horizon coincides with $\mathcal{H}$, it behaves like a $2+1$ dimensional dissipative fluid membrane with its energy-momentum tensor given by the following relation \cite{PhysRevD.58.064011, Maggio:2020jml,Chatterjee:2010gp, PhysRevD.95.064036}
\begin{equation}\label{ST_tensor}
\tau_{ab}=\rho u_a u_b+(p-\zeta \Theta)\gamma_{ab}-2\eta \sigma_{ab}~.
\end{equation}
Here $\rho$  and $p$ represents the density and pressure of the fluid while $\zeta$ and $\eta$ denotes the bulk and shear viscosity. $\Theta$ is the expansion, $\sigma_{ab}$ is the shear tensor, $u_a$ is the 3-velocity of the fluid and $\gamma_{ab}=h_{ab}+u_a u_b$ is the spacelike cross section of $\mathcal{S}$. For a black hole in general relativity, $\rho$ and $p$ depend on the spacetime geometry whereas the parameters $\zeta$ and $\eta$ take the following value $\zeta_{\textrm{BH}}\equiv -1/16\pi$ and $\eta_{\textrm{BH}}\equiv 1/16\pi$ respectively. \\
 In this paper, we are interested in studying the ringdown properties of a charged compact objects. Hence we consider that the fiducial observers are located at the surface of a charged compact object. The geometry outside the object is described by a \RN\ space-time \cite{Wald:1984rg,Chandrasekhar:1985kt,Hobson:2006se,Poisson:2009pwt},
\begin{equation}\label{rn_bh}
ds^{2}=-f(r)~dt^{2}+\frac{dr^{2}}{f(r)}+r^{2}(d\theta^{2}+\sin^{2}{\theta}~d\phi^{2})~,    
\end{equation}
which is a solution of the Einstein-Maxwell equation
\begin{eqnarray}\label{Ein_Max_Unpert}
&{G}_{\mu\nu}=8\pi T_{\mu\nu}~\,,\quad{F}=-\dfrac{Q}{r^2}~dt\wedge dr\\
 &T_{\mu\nu}=\dfrac{1}{4\pi}\bigg[{F}_{\mu\alpha}{F}_{\nu}^{~\alpha}-\dfrac{g_{\mu\nu}}{4}{F}_{\gamma\delta}{F}^{\gamma\delta}\bigg]~ ~\,,\quad{F^{\mu\nu}_{~~;\nu}}=0~.
\end{eqnarray}
Here, the function $f(r)=1-2M/r+Q^{2}/r^{2}$ has two zeros at $r_e=M+\sqrt{M^2-Q^2}$ and $r_c=M-\sqrt{M^2-Q^2}$, corresponding to the location of its event and Cauchy horizon. The surface of the compact object is situated at $R=r_e(1+\epsilon)$, where $\epsilon$ is a measure of the compactness of the object. In the limit $\epsilon \to 0$, the stretched horizon coincides with the event horizon. Using the condition given in \autoref{Membrane_paradigm} for the spacetime mentioned above, we obtain the following expression for $\rho$ and $p$ at the surface of the compact object \cite{Maggio:2020jml},
\begin{eqnarray}\label{rho_p}
&{\rho}(R)=-\dfrac{\sqrt{f(R)}}{4\pi R}\,,\qquad{p}(R)=\dfrac{2f(R)+f'(R)R}{16\pi R\sqrt{f(R)}}~
\end{eqnarray}
Note that the pressure term diverges, whereas the density vanishes at the black hole limit $\epsilon\to 0$.
\section{Perturbation of charged compact objects}\label{Sec_Perturb}
In this section, we discuss governing equations and the boundary conditions for the perturbations of charged compact objects. The perturbation equations for \RN\ black holes were obtained by several authors using different methods\cite{Moncrief:1975sb, PhysRevD.9.860,Chand}. Here, we mainly adopt the method described in \cite{PhysRevD.9.860}.\par  We consider the first-order perturbations $\delta g_{\mu\nu}$ and $\delta F_{\mu\nu}$ in the background quantities  ${g}_{\mu\nu}$ and ${F}_{\mu\nu}$ respectively so that the perturbed metric and field tensor becomes $\tilde{g}_{\mu\nu}={g}_{\mu\nu}+\delta g_{\mu\nu}$ and $\tilde{F}_{\mu\nu}=F_{\mu\nu}+\delta F_{\mu\nu}$. By expanding  $\delta g_{\mu\nu}$ and $\delta F_{\mu\nu}$ in terms of tensor harmonics and making use of the diffeomorphism symmetry of the system, we find that the only non-vanishing components of $\delta g_{\mu\nu}$ in the odd parity sector (in Regge-Wheeler gauge) are \cite{PhysRevD.9.860},   
 \begin{equation}\label{perturb_tensors}
 \begin{aligned}
 &\delta g_{t\phi}=h_0(r)~e^{i\omega t}~ \sin\theta ~\partial_\theta Y_{l0}\\
 &\delta g_{r\phi}=h_1(r)~e^{i\omega t}~  \sin\theta ~\partial_\theta Y_{l0}\\
 \end{aligned}  
 \end{equation}
 and their symmetric counterparts. Here, we set the azimuth quantum number $m$ to zero without losing any generality. Similarly, the only independent component of the perturbed field tensor is $\delta F_{\theta\phi}$. Substituting the expressions for $\delta g_{t\phi}$, $\delta g_{r\phi}$ and $\delta F_{\theta\phi}$ in Einstein-Maxwell equation \autoref{Ein_Max_Unpert}, it can be shown that the functions $h_0$ and $h_1$  are related by the following expression, $h_0=(if(r)/\omega)\partial_r(h_1 f(r))$ \cite{PhysRevD.9.860}. Redefining the function $h_1(r)=(2 i \omega r U(r))/\mu f(r)$ and $\delta F_{\theta\phi}=l(l+1) H(r) $, we obtain a pair for coupled differential equations for $U(r)$ and $H(r)$ where $\mu^2=l(l+1)-2$\cite{Chandrasekhar:1985kt}. 
 The equations for $U(r)$ and $H(r)$ can be decoupled if we use the following substitution \cite{Chandrasekhar:1985kt}, 
 \begin{equation}\label{Z1_Z2}
 \begin{aligned}
 &{Z_1}=q_1 H(r)+\sqrt{-q_1 q_2}U(r)\\
 &{Z_2}=-\sqrt{-q_1 q_2} H(r)+q_1 U(r)\\
 \end{aligned}  
 \end{equation}
 
 where the expresssion for $q_1$ and $q_2$ can be written as follows,
 \begin{equation}\label{q1_q2}
 \begin{aligned}
 &{q_1}=3 M+\sqrt{9 M^2+4 \left(l^2+l-2\right) Q^2}\\
 &{q_2}=3 M-\sqrt{9 M^2+4 \left(l^2+l-2\right) Q^2}\\
 \end{aligned}  
 \end{equation}
 The odd parity (axial) perturbation equation for a charged compact object takes the following form \cite{Chandrasekhar:1985kt}
\begin{equation}\label{perturb_eq}
\frac{d^2 Z_{i}}{dr_{*}^2}+\big[\omega^2-V_{i}(r)\big]Z_{i}=0  
\end{equation}
where $dr_{*}=dr/f(r)$ is the tortoise co-ordinate, 
\begin{equation}\label{perturb_potential}
	\begin{aligned}
	&{V}_{i}(r)=\dfrac{f(r)}{r^2} \left[l (l+1)+\dfrac{4 Q^2}{r^2}-\dfrac{q_j}{r}\right]\\
	\end{aligned}
\end{equation}
and $i,j=1,2$ ($i\neq j$). Note that, in the $Q\to 0$ limit, perturbation equation for $Z_{2}$ corresponds to Regge-Wheeler equation \cite{PhysRev.108.1063,Chandrasekhar:1985kt}. \par
\begin{table}[]
	\centering
	\def\arraystretch{1.2}      	
	\setlength{\tabcolsep}{1.4em}
	\begin{tabular}{|c|c|} 
		\hline
		\text{Q} & $\omega _{\text{BH}}$ \\
		\hline
		0.0 & 0.373673895745753-0.088969864662434 i \\
		\hline
		0.4 & 0.378117882893637-0.089365820010727 i \\
		\hline
		0.6 & 0.385757219512767-0.089607106118930 i \\
		\hline
		0.8 & 0.400503177075279-0.088609031463709 i \\
		\hline
	\end{tabular}
	\caption{The quasinormal frequencies of a \RN\ black hole in the fundamental $l=2$ mode. }\label{table3}
\end{table} 
To study these compact objects' ringdown properties, we need to impose the boundary conditions at infinity and the surface of the object. In the case of a black hole, quasinormal modes are defined as the solution of \autoref{perturb_eq} with the following boundary condition that there are only incoming waves at the event horizon outgoing waves at the infinity. The quasinormal modes of a \RN\ black hole in fundamental $l=2$ mode are presented in \autoref{table3} for different values of charge parameter $Q$. The outgoing boundary condition remains true for compact objects. However, the boundary conditions at the surface of the object get modified according to the reflectivity and compactness of the object. In this paper, we rely on the membrane paradigm to obtain generic boundary conditions on the compact object's surface $\mathcal{S}$. Details of the calculation are presented in  \autoref{App_boundary_condi}. Here, we briefly outline the necessary steps to obtain the boundary conditions. Up to the first order of perturbation in the axial sector, the normal vector to the stretched horizon $n^\mu$ remains unaltered. This indicates that the only non-vanishing components of the extrinsic curvature are $\delta K_{03}$ and $\delta K_{23}$ and their symmetric counterparts (see \autoref{App_boundary_condi} for further discussion). Moreover, the metric component $\delta g_{t\phi}$ imparts rotation to the compact object. Consequently, the fluid at the membrane now has a velocity component $\delta u^\phi$ along the azimuth direction while its other velocity components remain unchanged. By replacing the expression for 
$\delta K_{03}$, $\delta K_{23}$ and $\delta u^\phi$ in \autoref{Membrane_paradigm} and \autoref{ST_tensor}, we obtain the boundary condition for $U(r)$ at the surface of the compact object as given by the following expression (see \autoref{App_boundary_condi})
\begin{equation}\label{grav_pert}
\begin{aligned}
&U'\left(r_*\right)\bigg{|}_{r_*(R)}=\bigg[\frac{\mu  Q f }{P_h}H(r)\\&-\left(\frac{f \left[r \left(\left(\mu ^2+2\right) r-6 M\right)+4 Q^2\right]}{2 r P_h}+\frac{i \omega }{16 \pi  \eta }\right)U\left(r\right)\bigg]_{r=R}
\end{aligned}
\end{equation}
where $\mu^2=l(l+1)-2$, $P_h=r (r-3 M)+2 Q^2$ and $R$ is the radius of the compact object. \\
 In order to obtain the boundary condition for $H(r)$, we notice that the membrane paradigm predicts the existence of surface 4-current $j^{\mu}_s$ along the surface of the object \cite{Thorne, PhysRevD.58.064011}. The surface current $j^{\mu}_s$ is related to the field tensor $\tilde{F}^{\mu\nu}$ by the following expression,
\begin{equation}
j^{\mu}_{s}=\frac{1}{4\pi}\bigg(\tilde{F}^{\mu\nu}n_{\nu}\bigg)\,.\label{surface_current}    
\end{equation}
The above expression can be written in terms of the electric and magnetic field observed by the FIDOs as follows,
\begin{equation}\label{surface_EM_field}
\begin{aligned}
	&{E}^\bot_{\textrm{FIDO}}={4\pi j^{0}_s}=4\pi \sigma_{e} \\&(\vec{B}^\parallel_{\textrm{FIDO}})^A=4\pi (\vec{j}_s\times \hat{n})^{A}~.
\end{aligned}
\end{equation}
\autoref{surface_EM_field} tells that the normal component of the electric field and the tangential component magnetic field is discontinuous at the membrane due to the presence of surface charge and current, respectively, following the laws of electrodynamics.
To determine the boundary condition on the compact object's surface, we consider a set of radially freely falling observers (FFO) who satisfies the following equation of motion, $dr/dt=-f\sqrt{1-f}$. To them, the fiducial observers sitting on the surface of the compact object (FIDO) are moving outward with velocity \cite{Thorne} 
\begin{equation}\label{FIDO}
v=-\dfrac{1}{\sqrt{f}}\dfrac{dr}{d\tau}\bigg|_{r=R}=\sqrt{1-f(R)}~,
\end{equation}
where $\tau$ is the time measured by the FIDO at $\mathcal{S}$ (see \autoref{App_RN_perturb}). Note that the fields measured by FFOs are finite. However, FIDO measurements are Lorentz boosted as follows \cite{Thorne},
\begin{eqnarray}\label{lorentz_boost}
	&{E}^r_{\textrm{FIDO}}=E^r_\textrm{FFO} \,,\qquad{B}^r_{\textrm{FIDO}}=B^r_{\textrm{FFO}}\nonumber~,\\
	&{E}^\theta_{\textrm{FIDO}}=\gamma\left(E^\theta_{\textrm{FFO}}-v B^\phi_{\textrm{FFO}}\right) \nonumber~,\\&{B}^\phi_{\textrm{FIDO}}=\gamma\left(B^\phi_{\textrm{FFO}}-v {E}^\theta_{\textrm{FFO}}\right)~,\\
	&{E}^\phi_{\textrm{FIDO}}=\gamma\left(E^\phi_{\textrm{FFO}}+v B^\theta_{\textrm{FFO}}\right)\nonumber~, \\&{B}^\theta_{\textrm{FIDO}}=\gamma\left(B^\theta_{\textrm{FFO}}+v {E}^\phi_{\textrm{FFO}}\right)\nonumber~.
\end{eqnarray}
where $\gamma=1/\sqrt{1-v^2}=1/\sqrt{f}$ is the Lorentz factor. Now consider the following term, $(\hat{n}\times \vec{B}^\parallel_{\textrm{FIDO}})^{\theta}=-B^{\phi}_{\textrm{FIDO}}$. For a compact enough object, $f(r)$ is small on the surface object, hence we can expand $v$ as $v=1-(1/2)f+\mathcal{O}(f^2)$. Under such consideration, $(\hat{n}\times \vec{B}^\parallel_{\textrm{FIDO}})^{\theta}$ becomes
\begin{equation}\label{theta_fido}
	\begin{aligned}
	&(\hat{n}\times B^\parallel_{\textrm{FIDO}})^{\theta}=-B^{\phi}_{\textrm{FIDO}}\\&=\gamma\Big[(1-(1/2)f) {E}^\theta_{\textrm{FFO}}-v(1+(1/2)f)B^\phi_{\textrm{FFO}}\Big]+\mathcal{O}(f^\frac{3}{2})\\
	&=\gamma\left(E^\theta_{\textrm{FFO}}-v B^\phi_{\textrm{FFO}}\right)-\dfrac{f}{2}\gamma\left(E^\theta_{\textrm{FFO}}+ B^\phi_{\textrm{FFO}}\right)+\mathcal{O}(f^\frac{3}{2})\\
	&={E}^\theta_{\textrm{FIDO}}-{E}^\theta_{\textrm{Cor}}
	\end{aligned}
\end{equation}
where, ${E}^\theta_{\textrm{Cor}}=\gamma(f/2)\left(E^\theta_{\textrm{FFO}}+ B^\phi_{\textrm{FFO}}\right)+\mathcal{O}(f^\frac{3}{2})$ is the correction term. Here, in the first line we have used the following relation $v(1+(1/2)f+\mathcal{O}(f^2))\approx1$.  Similarly, we can show that $(\hat{n}\times \vec{B}^\parallel_{\textrm{FIDO}})^{\phi}$ can be written as follows,
\begin{equation}\label{phi_fido}
	\begin{aligned}
		(\hat{n}\times \vec{B}^\parallel_{\textrm{FIDO}})^{\phi}=B^{\theta}_{\textrm{FIDO}}&={E}^\phi_{\textrm{FIDO}}-{E}^\phi_{\textrm{Cor}}
	\end{aligned}
\end{equation}
where, ${E}^\phi_{\textrm{Cor}}=\gamma(f/2)\left(E^\phi_{\textrm{FFO}}- B^\theta_{\textrm{FFO}}\right)+\mathcal{O}(f^\frac{3}{2})$.  \autoref{theta_fido} and \autoref{phi_fido} can be written in a more compact form as follows,
\begin{equation}\label{parallel_fields}
	\begin{aligned}
		\vec{E}^\parallel_{\textrm{FIDO}}=(\hat{n}\times \vec{B}^\parallel_{\textrm{FIDO}})+\vec{E}^\parallel_{\textrm{Cor}}~.
	\end{aligned}
\end{equation}
Note that, since $\gamma=1/\sqrt{f}$, the main contribution for $(\hat{n}\times \vec{B}^\parallel_{\textrm{FIDO}})$ comes from the first term of \autoref{theta_fido} and \autoref{phi_fido} for a compact enough object. Moreover, in the black hole limit $f\to 0$, $\vec{E}^\parallel_{\textrm{Cor}}$ vanishes identically. The physical reasoning behind this is as follows, in the black hole limit FIDOs are moving outward at near the speed of light (see \autoref{FIDO}) with respect to any timelike observer and thus they see the tangential component of the electromagnetic fields as incoming waves \cite{Thorne, PhysRevD.58.064011}. For axial perturbation, the only non-vanishing component of $\vec{E}^\parallel_{\textrm{FIDO}}$ is $\vec{E}^\phi_{\textrm{FIDO}}$
\begin{equation}\label{E_phi}
	\begin{aligned}
		{E}^\phi_{\textrm{FIDO}}&=(\hat{n}\times \vec{B}^\parallel_{\textrm{FIDO}})^{\phi}+{E}^\phi_{\textrm{Cor}}\\
		&=-4\pi \tilde{F}^{r\phi}n_r+{E}^\phi_{\textrm{Cor}}~.
	\end{aligned}
\end{equation}
where we have used \autoref{surface_EM_field} in the second line. Moreover, the surface current on the stretched horizon along the azimuth direction can be written as follows, $j^\phi=\sigma_{e} \delta u^\phi$. The expression for resistivity $\rho_{s}$ can be obtained via Ohm's law,
\begin{equation}\label{resistivity_1}
	\begin{aligned}
		\rho_{s}=\dfrac{{E}^\phi_{\textrm{FIDO}}}{j^\phi}=\dfrac{-4\pi \tilde{F}^{r\phi}n_r}{\sigma_{e} \delta u^\phi}+\rho_{\textrm{Cor}}~,
	\end{aligned}
\end{equation}
where $\rho_{\textrm{Cor}}={E}^\phi_{\textrm{Cor}}/j^\phi$. Note that, the expression for $\rho_{\textrm{Cor}}$ depends on parameters of the compact object as well as its compactness. Moreover, in the black hole limit, this term vanishes identically and $\rho_{s}$ becomes $\rho_{s}=\rho_{\textrm{SBH}}\equiv 4\pi$. By redefining the resistivity as $\rho_{S}=\rho_{s}-\rho_{\textrm{Cor}}$, we can express the above equation as follows,
\begin{equation}\label{resistivity_2}
	\begin{aligned}
		\rho_{S}=\dfrac{-4\pi \tilde{F}^{r\phi}n_r}{\sigma_{e} \delta u^\phi}~.
	\end{aligned}
\end{equation}
Substituting the expression for $\tilde{F}^{r\phi}$, $n_r$, $\sigma_{e}$ and $\delta u^\phi$ in \autoref{resistivity_2}, we obtain the boundary condition for $ H(r) $ (or, equivalently for $\delta F_{\theta\phi}$  ) at the surface of the compact object as follows,
\begin{equation}\label{em_pert}
\begin{aligned}
&H'\left(r_*\right)\bigg{|}_{r_*(R)}=\Bigg[-\frac{2 Q^2 f(r) H\left(r\right)}{r P_h(r)}\\&+U\left(r\right) \left(\frac{\mu  Q f(r)}{P_h(r)}+\frac{i Q \omega  \left(4 \pi -\rho _S\right)}{32 \pi ^2 \eta  \mu  r}\right)\Bigg]_{r=R}
\end{aligned}
\end{equation}
\begin{figure*}[th]
	\centering
		\begin{minipage}[b]{0.48\textwidth}
			\includegraphics[width=\textwidth]{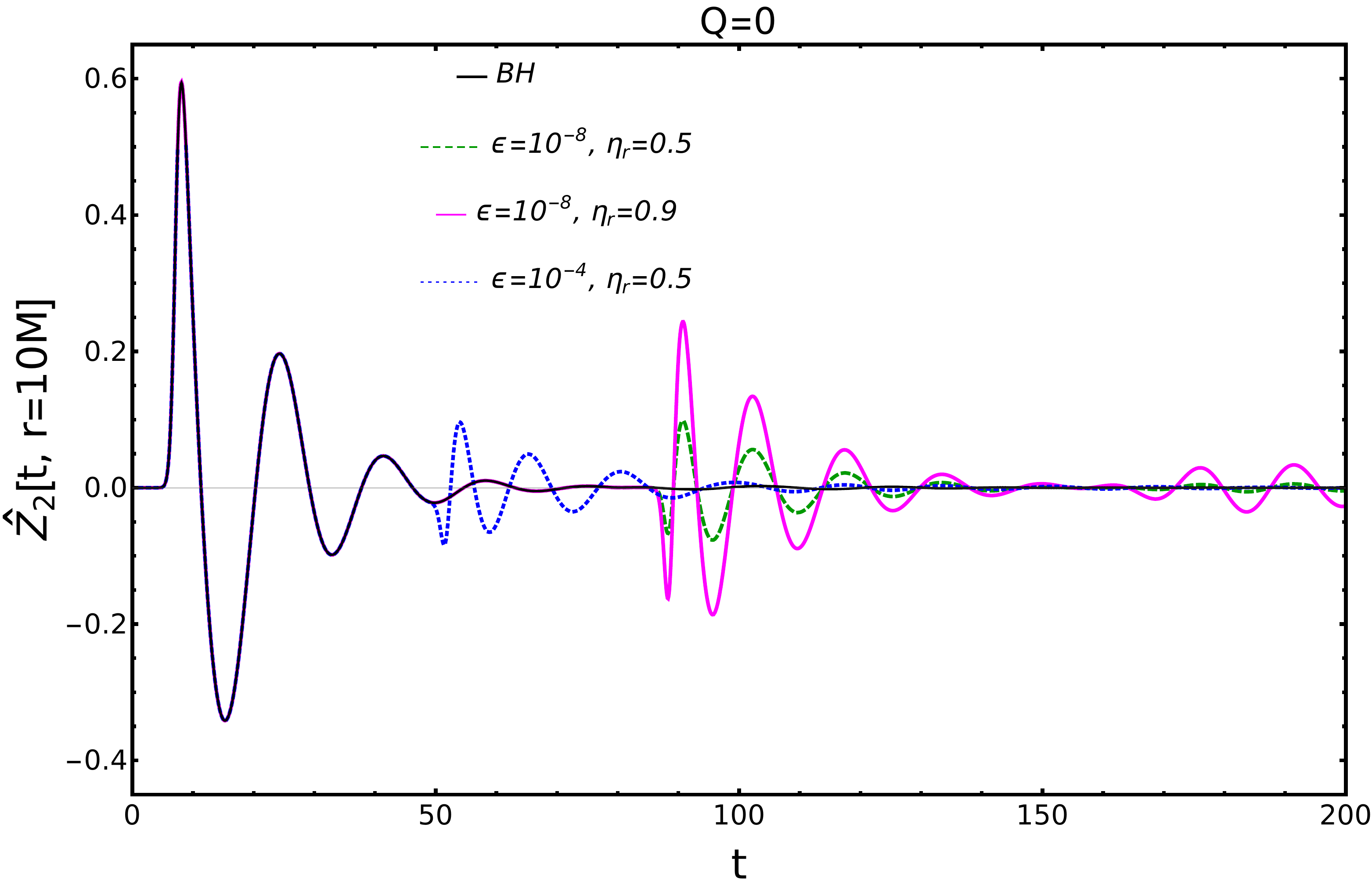}
		\end{minipage}
		\hfill
		\begin{minipage}[b]{0.48\textwidth}
			\includegraphics[width=\textwidth]{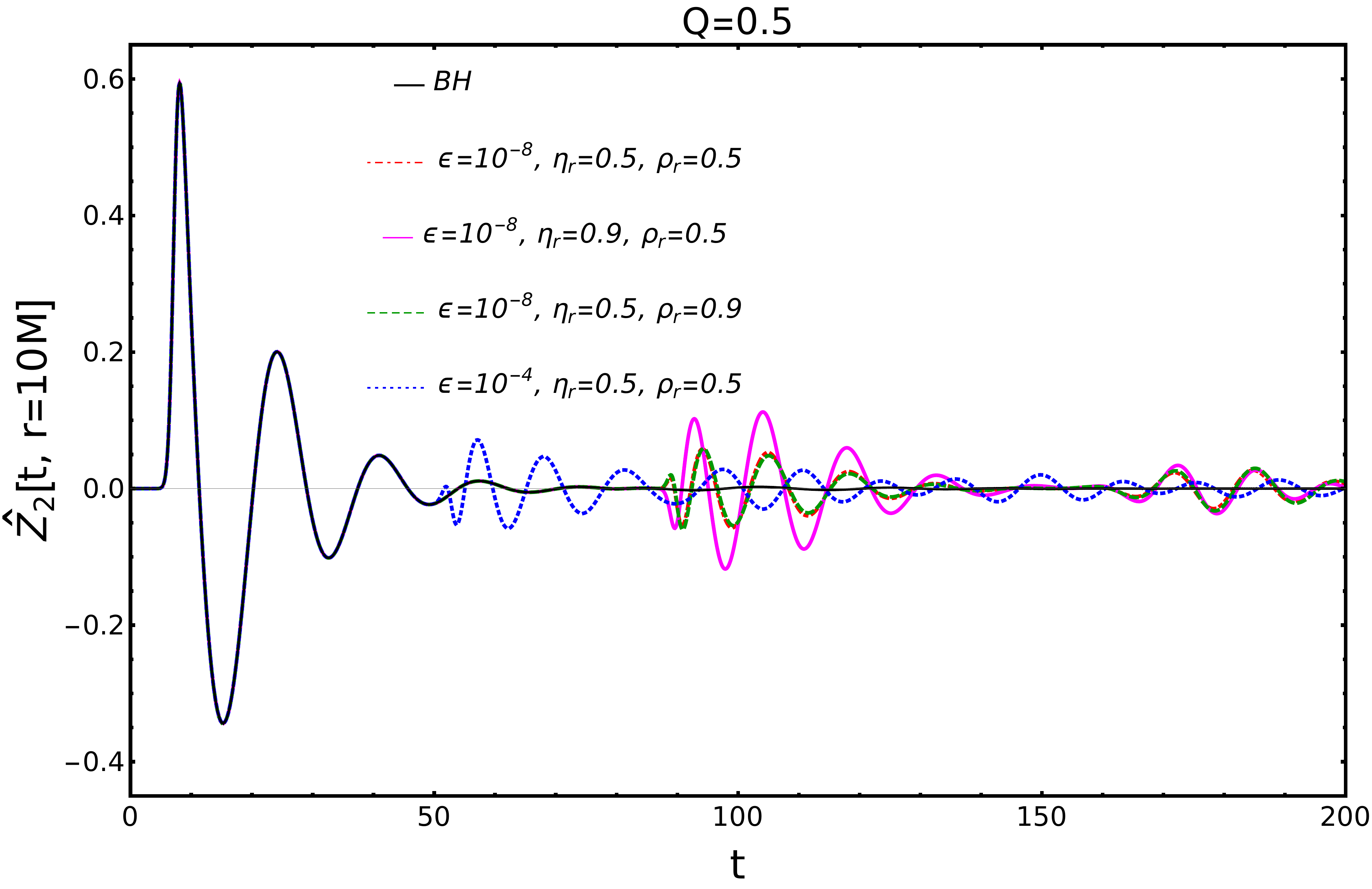}
		\end{minipage}\par
			\hfill
			\begin{minipage}[b]{0.48\textwidth}
				\includegraphics[width=\textwidth]{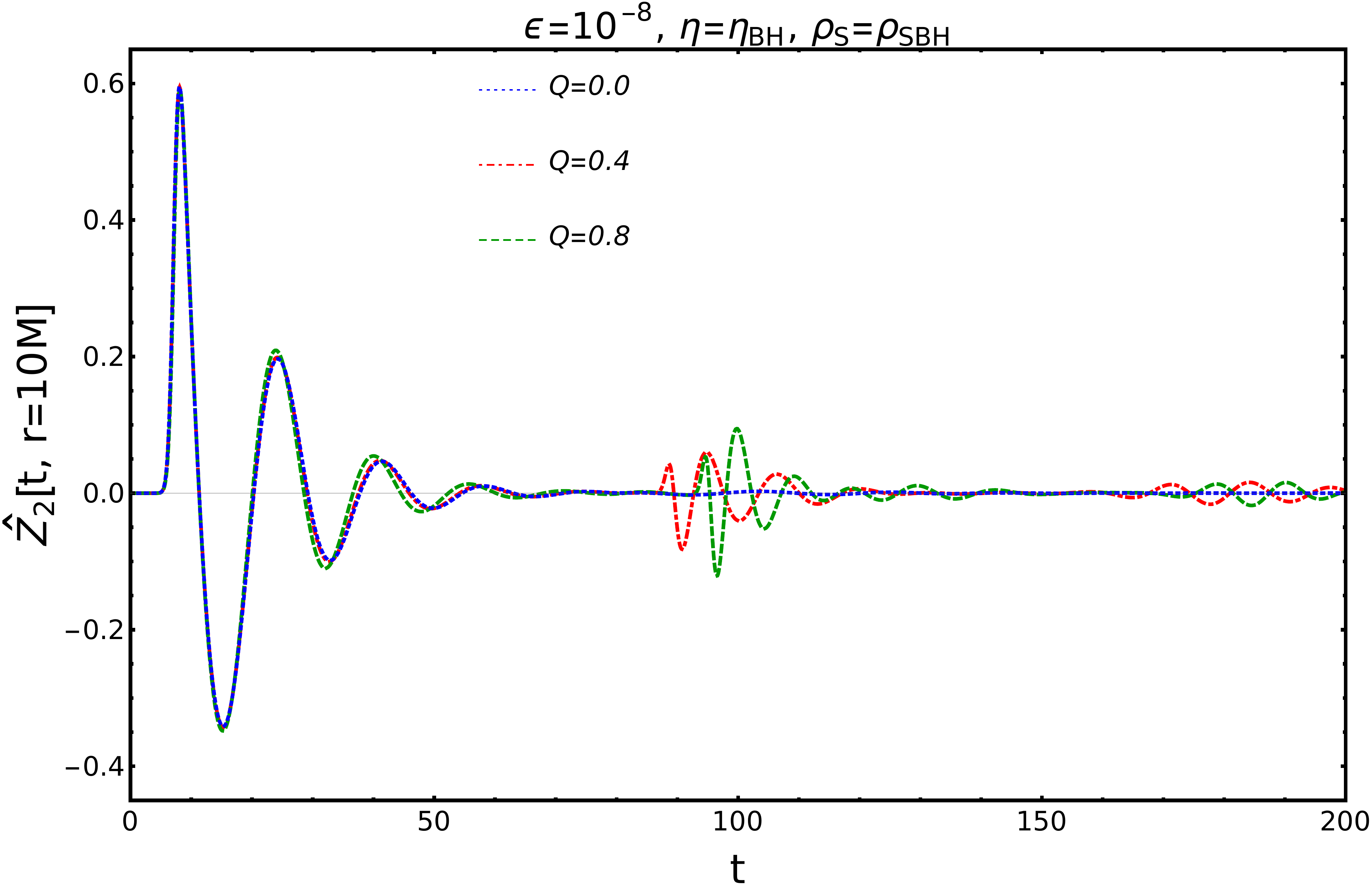}
			\end{minipage}
				\hfill
				\begin{minipage}[b]{0.48\textwidth}
					\includegraphics[width=\textwidth]{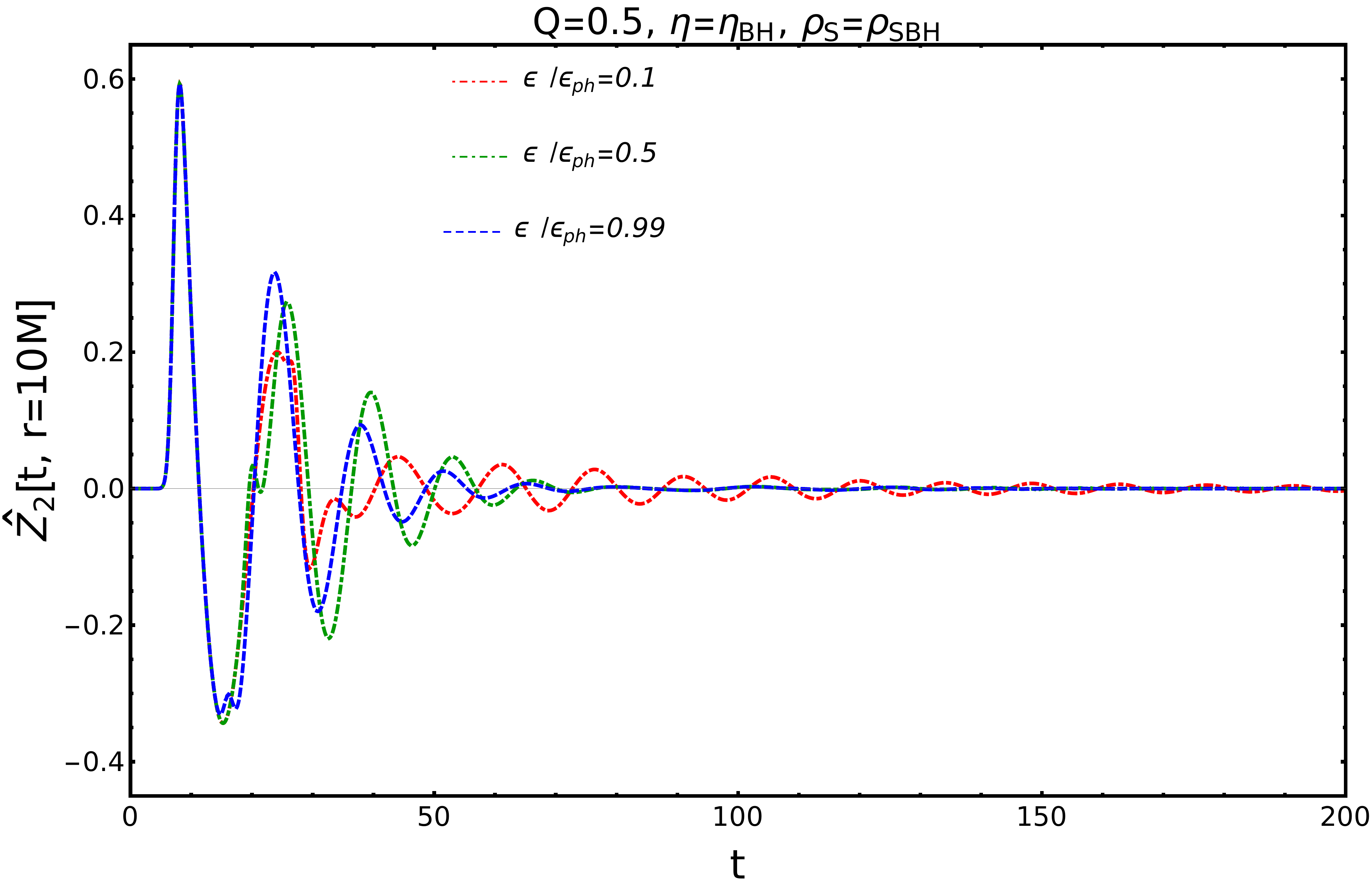}
				\end{minipage}
	\caption{A comparative study of the ringdown properties of differnt charged compact objects with different values charge $Q$, compactness parameter $\epsilon$, shear viscosity $\eta$ and resistivity $\rho_{S}$ is presented for $l=2$ mode.
		Here, $\eta_r\equiv1-\eta/\eta_{\textrm{BH}}$, $\rho_r\equiv1-\rho_{S}/\rho_{\textrm{SBH}}$ and $\epsilon_{\textrm{ph}}$ is the value of compactness parameter when the surface of the compact object coincides with its photon sphere. In the top panels, the ringdown waveforms of highly compact ($\epsilon\lesssim 0.0001$) neutral (in the left) and charged (in the right) objects are presented. In this scenario, GW echoes are present. In each of these plots, the black curve represents the ringdown properties of black holes.  The bottom left panel shows the ringdown properties for different values of $Q$, whereas the bottom right panel depicts the same for relatively less compact objects ($\epsilon_{\textrm{ph}}>\epsilon\gtrsim 0.1$). As evident, the GW echoes are absent for higher values of $\epsilon$. }\label{Ringdown}
\end{figure*}
 Using the transformations outlined in \autoref{Z1_Z2}, we can rewrite the boundary conditions \autoref{grav_pert} and \autoref{em_pert} in terms of $Z_1$ and $Z_2$ which can be expressed as follows,
\begin{widetext}
\begin{equation}\label{bc1}
\begin{aligned}
{\frac{d Z_{1}}{dr_{*}}}\bigg{|}_{r_*(R)}&=\Bigg[-\frac{1}{D}\Bigg(\frac{i Q^2 \omega\left(4 \pi  \mu ^2 r-q_1 (\text{$\rho_{S} $}-4 \pi )\right)}{16 \pi ^2 \eta  r}+\dfrac{2 Q^2 f(r) \bigg(6 M q_1+\mu ^2 \bigg((l(l+1) r+6 M)r-2 q_2 r+8 Q^2\bigg)\bigg)}{r P_h}\Bigg) Z_{1}\\
&-\frac{1}{D}\Bigg(\dfrac{-\frac{1}{2} i q_1 Q\omega \left(q_1 \left(4 \pi -\rho _S\right)+4 \pi  \mu ^2 r\right)}{16 \mu  r \pi ^2 \eta}-\dfrac{ \pi ^2 r\eta  \mu  l(l+1) q_1 Q f(r)}{P_h}\Bigg)Z_{2}\Bigg]_{r=R}
\end{aligned}
\end{equation}

\begin{equation}\label{bc2}
\begin{aligned}
{\frac{d Z_{2}}{dr_{*}}}\bigg{|}_{r_*(R)}&=\Bigg[-\dfrac{1}{8r D}\Bigg(\frac{i \mu  Q \omega\left(Q^2 \left(4 \pi -\rho _S\right)-\pi  q_1 r\right)}{\pi ^2 \eta }-\frac{8 \mu  l(l+1) q_1 Q r^2 f(r)}{P_h}\Bigg) Z_{1}\\
&-\frac{1}{D}\Bigg(\frac{f(r) \left(2 \mu ^2 Q^2 \left(l (l+1) r^2-18 M r+2 q_2 r+8 Q^2\right)+3 M q_1 \left(r (l (l+1) r-6 M)+4 Q^2\right)\right)}{r P_h}\\
&+\frac{i\omega \left(q_1 \left(6 \pi  M r+Q^2 \rho _S-4 \pi  Q^2\right)+4 \pi  \mu ^2 Q^2 r\right)}{16 \pi ^2 \eta  r}\Bigg)Z_{2}\Bigg]_{r=R}
\end{aligned}
\end{equation}
\end{widetext}
where  $D=6 M q_1+8 \mu ^2 Q^2$. Note that, when the surface of the compact object coincides with the radius of the photon sphere $r_{\textrm{ph}}$, the term $P_h$ vanishes and the coefficient of $Z_1$ and $Z_2$ diverges in \autoref{bc1} and \autoref{bc2}. This implies $Z_1=Z_2=0$ in this limit regardless the value of the parameters $Q$, $\eta$, $\rho_{S}$. The same is also true when shear viscosity $\eta$ of the compact object vanishes.  
\begin{figure*}[t]
	\centering
	\minipage{0.48\textwidth}
	\includegraphics[width=\linewidth]{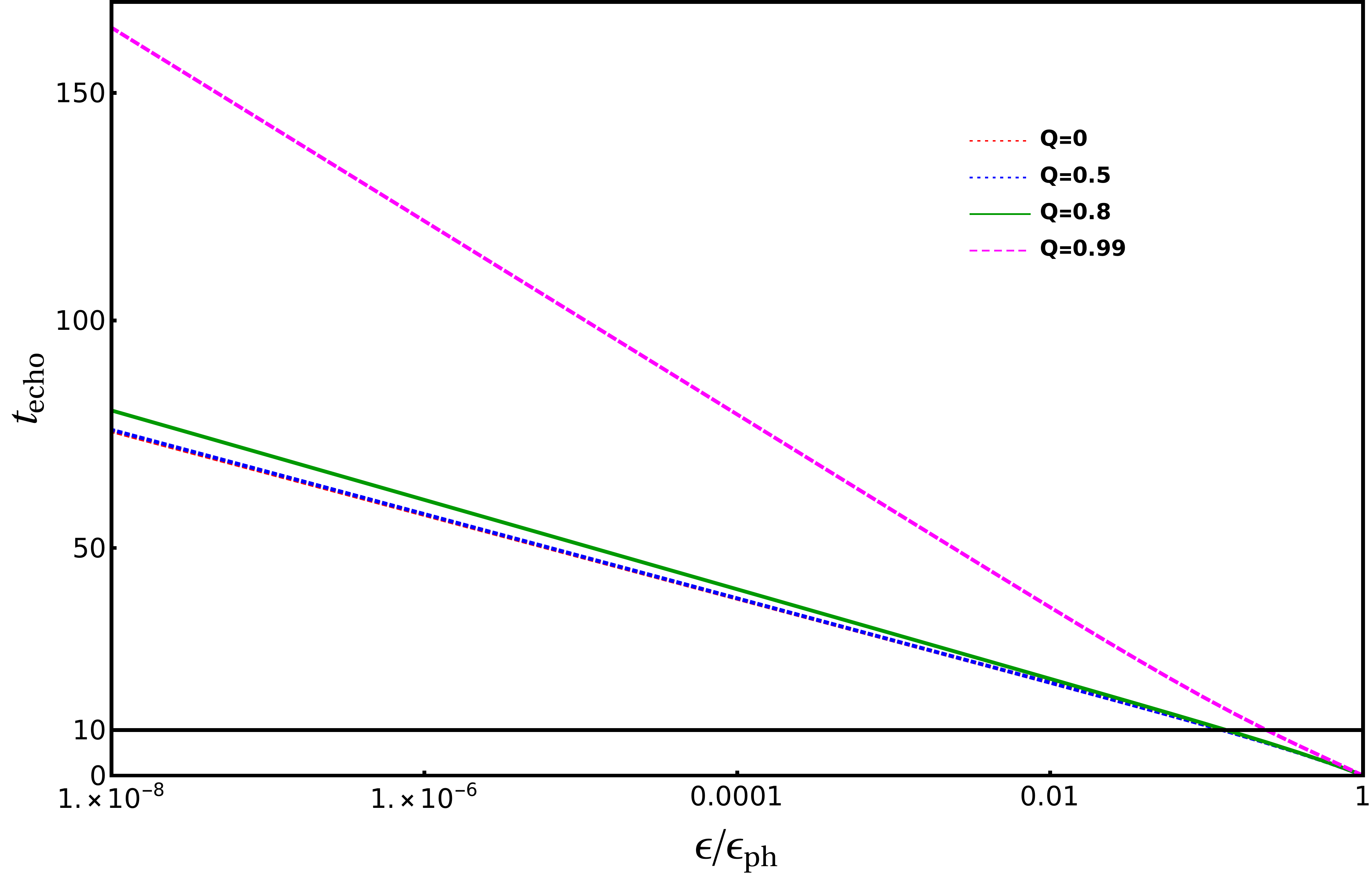}
	\endminipage\hfill
	\minipage{0.48\textwidth}
	\includegraphics[width=\linewidth]{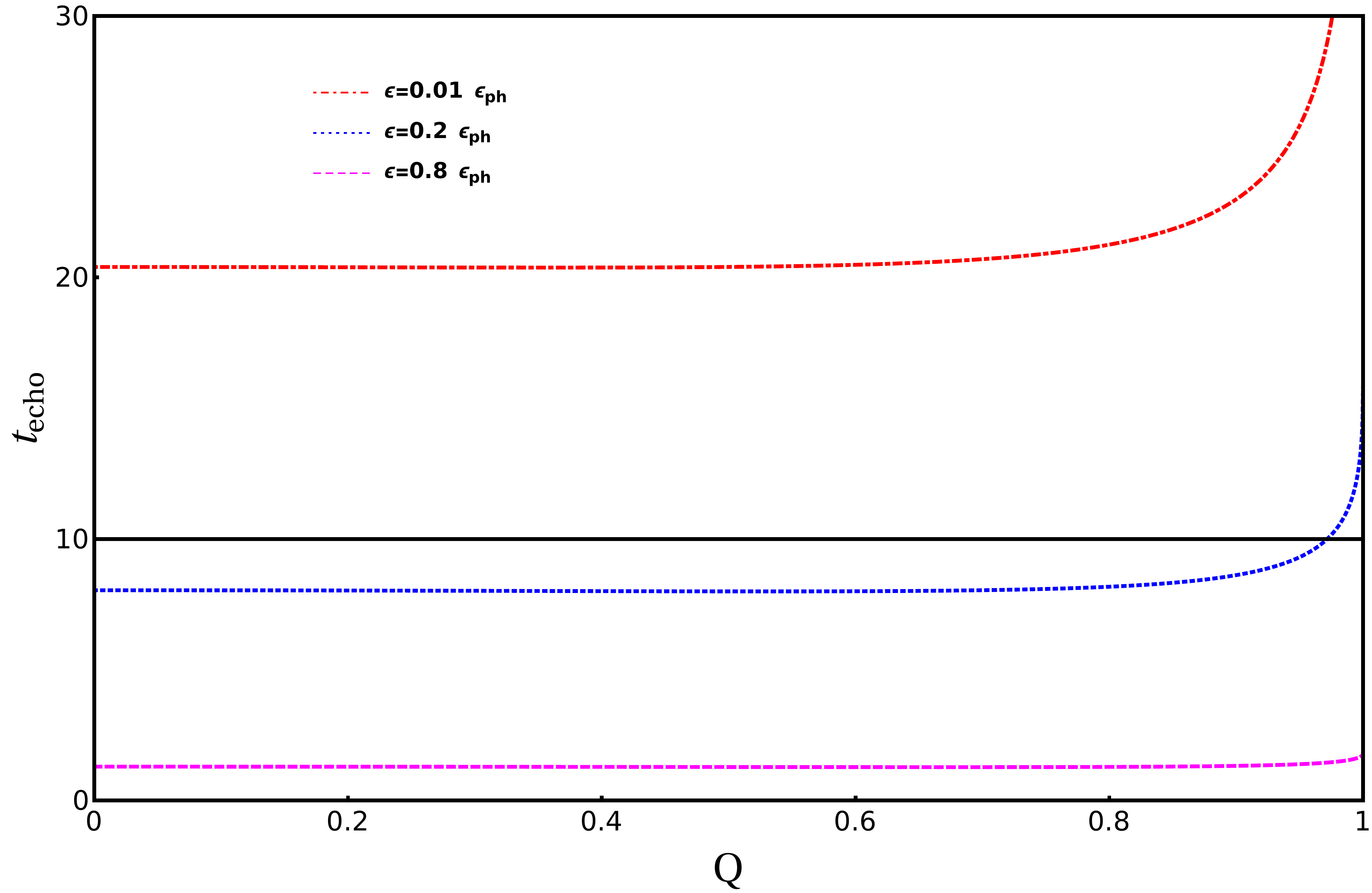}
	\endminipage
	\caption{ The variation of echo time $t_{\textrm{echo}}/M$ as a function of normalized compactness parameter $\epsilon/\epsilon_{\textrm{ph}}$ (left panel) and charge $Q$ (right panel) is plotted. In each of these plots, the solid black line represents the typical decay time scale of \RN\ black holes, $t_{\textrm{d}}= 1/\omega_{\textrm{I}}\approx 10 M$. As evident, the echo time decreases with the increase of $\epsilon$. The QNMs of the object get modified if $t_{\textrm{d}}>t_{\textrm{echo}}$. Clearly, this criteria is satisfied only for  the objects with $\epsilon\gtrsim 0.05\epsilon_{\textrm{ph}}$.}\label{echo_time}
\end{figure*}

\section{Ringdown of Charged Compact Objects}\label{Sec_Ringdown}

In this section, we discuss the ringdown properties of charged compact objects. It is quite well-known that when an astrophysical object gets perturbed, it emits gravitational waves to settle down to a stationary state. According to perturbation theory in static spacetime, the ringdown phase can be described in terms of the superposition of damped sinusoidal waves of the following form $h(t)=\sum_{nl}\mathcal{A}_{nl}~e^{-i\omega_{nl}t}Y_{nl}(\Omega)$ 
where $n$ is the overtone number, $Y_{nl}(\Omega)$ is the generalized Legendre polynomial and $\omega_{nl}$ are the complex frequencies known as the quasinormal modes (QNMs) whose real part $\omega_{\textrm{R}}$ denotes the frequency of the oscillation. In contrast, the imaginary part $\omega_{\textrm{I}}$ denotes the damping rate \cite{Vishveshwara:1970zz,Chandrasekhar:1975zza,Kokkotas:1999bd,Berti:2009kk}. These QNMs can be obtained by solving the perturbation equations of the form of \autoref{perturb_eq} with appropriate boundary conditions. In particular, there are only incoming waves at the event horizon in the black hole scenario. Moreover, these QNMs are characterized by the mass ($ M $), charge ($ Q $) and angular momentum ($ a $) of the black hole in accordance to the \textit{no-hair} theorem \cite{https://doi.org/10.1002/asna.19752960110,PhysRev.164.1776,PhysRevLett.26.331}. However, as we will see in the absence of the event horizon, these modes also depend on the reflection coefficient and the object's compactness. In order to study the time domain behavior these perturbed objects, we use an inverse Fourier transformation $-i\omega Z_{i}(\omega,r)\to \partial_{t}\hat{Z}_{i}(t,r_{*})$ in \autoref{perturb_eq}, \autoref{bc1} and \autoref{bc2} and solve these equations with the following initial conditions,
\begin{eqnarray}
	\hat{Z}_{i}(0,r_{*})=0\,,\qquad{\partial_{t}\hat{Z}_{i}(0,r_{*})}=e^{-(r_{*}-7)^2}~.
\end{eqnarray}
\begin{figure*}[th]
	\centering
	\minipage{0.48\textwidth}
	\includegraphics[width=\linewidth]{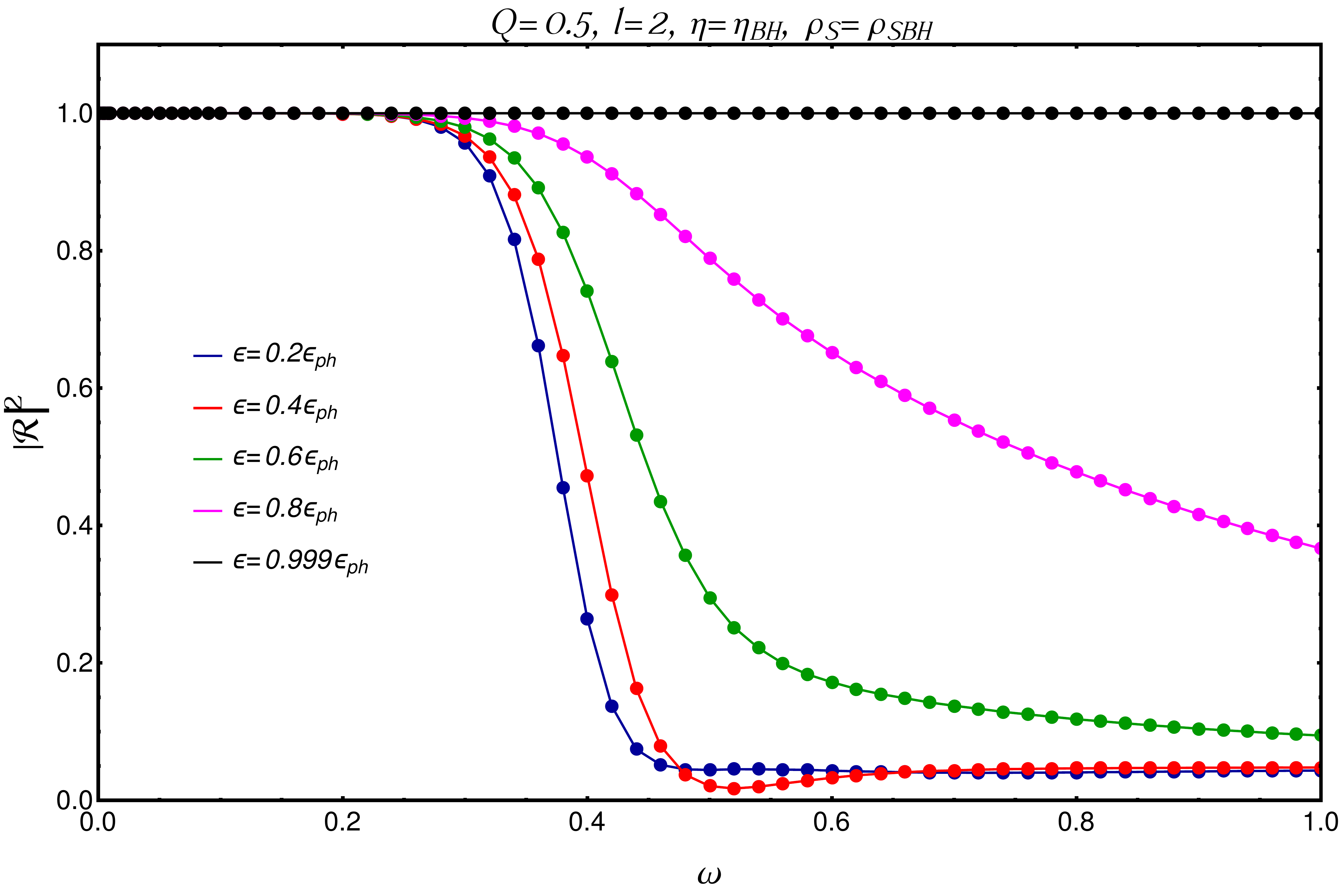}
	\endminipage\hfill
	\minipage{0.48\textwidth}
	\includegraphics[width=\linewidth]{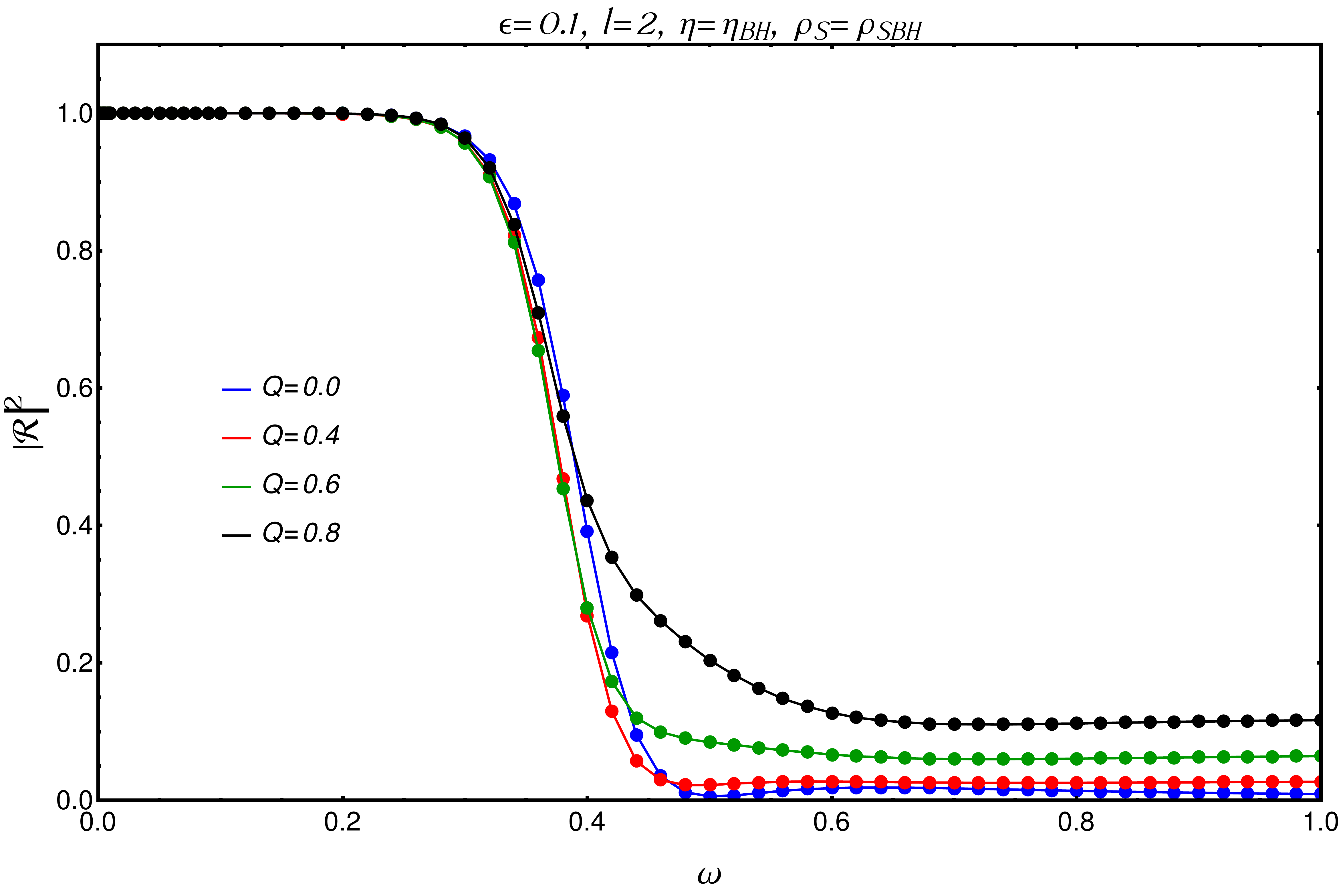}
	\endminipage\hfill
	\minipage{0.48\textwidth}
	\includegraphics[width=\linewidth]{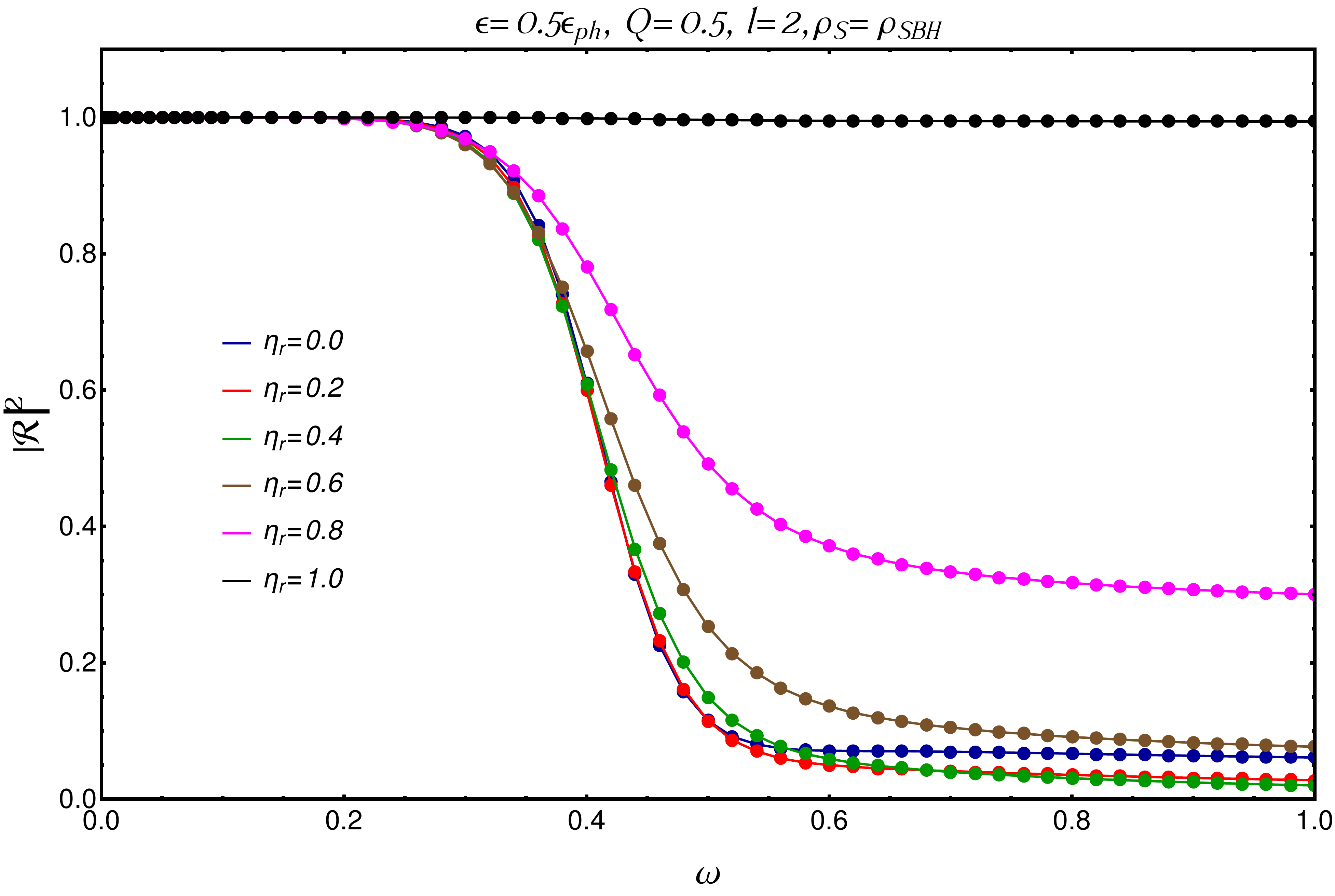}
	\endminipage\hfill
	\minipage{0.48\textwidth}
	\includegraphics[width=\linewidth]{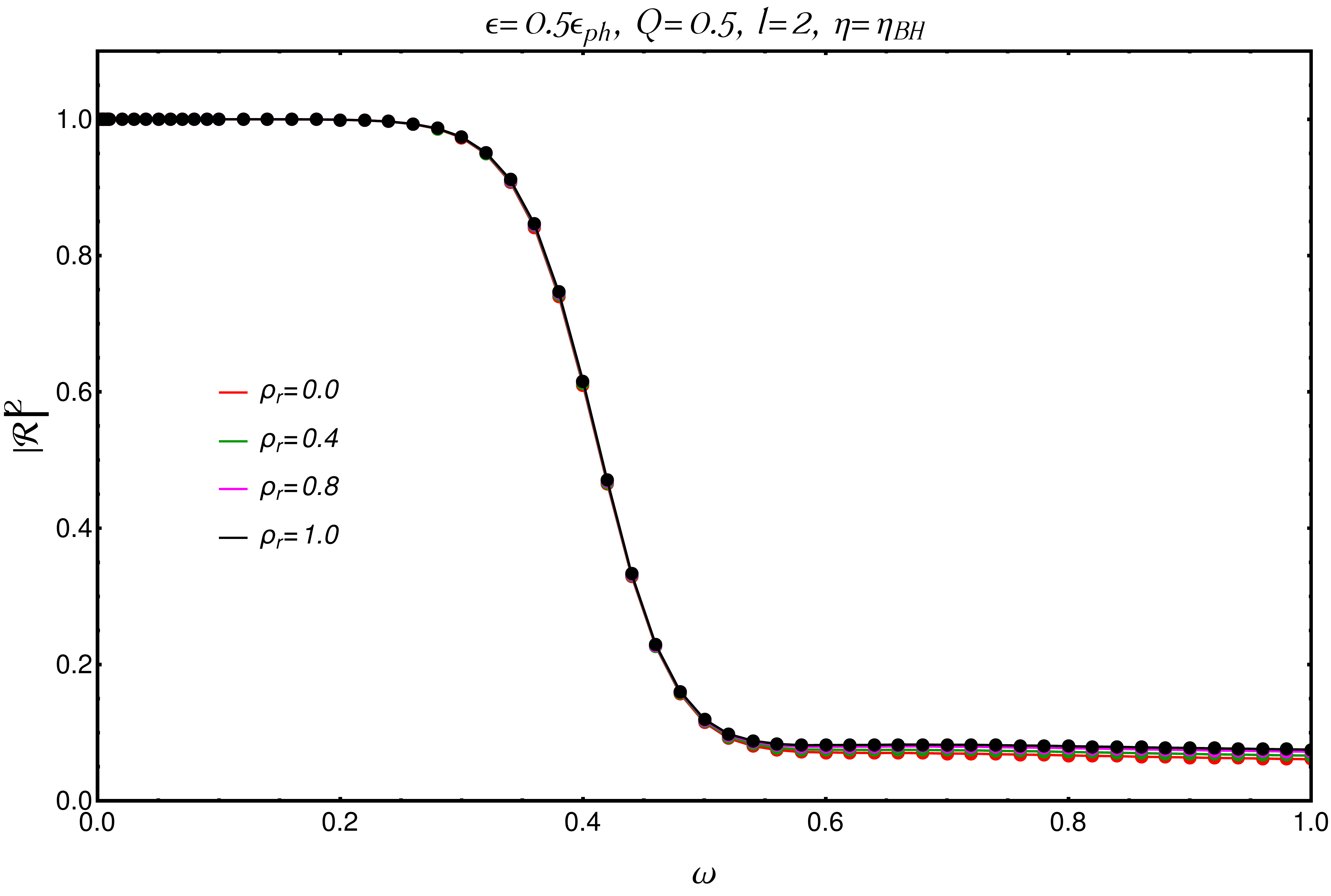}
	\endminipage
	\caption{The variation of reflection coefficient $|\mathcal{R}|^2$ as a function of $\omega$ for different values of compactness parameter $\epsilon$ (top left), charge $Q$ (top right), shear viscosity $\eta_r\equiv1-\eta/\eta_{\textrm{BH}}$ (bottom left) and resistivity $\rho_r\equiv1-\rho_{S}/\rho_{\textrm{SBH}}$ (bottom right) is presented. Here, $\epsilon_{\textrm{ph}}$ is the value of compactness parameter when the surface of the compact object coincides with its photon sphere.    }\label{Reflectivity}
\end{figure*}
In \autoref{Ringdown}, we plot the ringdown waveforms for different compact objects with different values of charge $Q$, compactness $\epsilon$, shear viscosity $\eta$ and resistivity $\rho_{S}$. In the subsequent discussions, we adopt the nomenclature provided in \cite{Cardoso:2019rvt,Cardoso:2017njb} to describe objects with different values of compactness parameter. Objects with compactness parameter $0.01<\epsilon<\epsilon_{\textrm{ph}}$ are termed as \textit{Ultra Compact objects} (UCOs) whereas objects with $\epsilon<0.01$ are dubbed as \textit{Clean-photon sphere objects} (ClePhOs) where $\epsilon_{\textrm{ph}}$ is the value of the compactness parameter when the radius of the object $R=r_e(1+\epsilon)$ coincides with the radius of the unstable photon orbit $r_{\textrm{ph}}$. The top panel of \autoref{Ringdown} shows the ringdown waveforms of neutral (left) and charged (right) ClePhOs for different values of $\eta_r\equiv1-\eta/\eta_{\textrm{BH}}$ and $\rho_r\equiv1-\rho_{S}/\rho_{\textrm{SBH}}$. In each of these plots, the black curves represent the ringdown of a black hole. As evident, the ringdown signals of ClePhOs are identical to those of the black holes in the beginning. However, we obtain repetitive signals in the form of \textit{gravitational wave echoes} from ClePhOs at later times whereas the signals die down for the black holes. Moreover, the time separation  $t_{\textrm{echo}}$ (so called the ``\textit{echo time}") between two consecutive signal decreases with the increase of $\epsilon$. \par The phenomenon can be explained through following arguments\cite{Cardoso:2019rvt,Cardoso:2016oxy,Cardoso:2017cqb,Maggio:2020jml}.
In geometrical-optics approximation, QNMs are interpreted in terms of photons trapped in unstable photon orbit $r_{\textrm{ph}}$ where the perturbation potential has its maxima. The real part of the frequency is related to the angular velocity of the photons at   $r_{\textrm{ph}}$. In contrast, the imaginary part is related to the decay time scale $t_{\textrm{d}}$ of null geodesics at the unstable photon orbit \cite{Cardoso:2008bp}. Thus, the radiations generated at $r_{\textrm{ph}}$ will reach the surface of the compact object in time. 
\begin{equation}\label{round_trip}
\begin{aligned}
\Delta t &=\int_{R}^{r_{\textrm{ph}}}\dfrac{dr}{f(r)}\\&=r_e(\epsilon_{\textrm{ph}}-\epsilon)-\frac{1}{2\kappa_{e}}\log[\frac{\epsilon}{\epsilon_{\textrm{ph}}}]+\frac{1}{2\kappa_{c}}\log[\frac{\epsilon+\varepsilon}{\epsilon_{\textrm{ph}}+\varepsilon}]~,
\end{aligned}
\end{equation}
where $\epsilon_{\textrm{ph}}=(r_{\textrm{ph}}-r_{e})/r_e$ and $\varepsilon=(r_{e}-r_{c})/r_e$. Part of radiations get reflected from the surface of object and give rise to the gravitational wave echoes  in time $t_{\textrm{echo}}=2\Delta t$ \cite{Abedi:2016hgu,Cardoso:2019rvt,Cardoso:2016oxy}. In the left panel of \autoref{echo_time}, we plot the variation of $t_{\textrm{echo}}$ with $\epsilon$ for different values of $Q$. As evident, $t_{\textrm{echo}}$ decreases with the increase of $\epsilon$ which is in accordance to the behavior of the gravitational wave echoes presented in the top panels of \autoref{Ringdown}. The effect of charge on the ringdown properties of ClePhOs is presented in the bottom left panel of \autoref{Ringdown}. As we can see that the  $t_{\textrm{echo}}$ increases  with the increase of $Q$. This phenomena can be explained by following the same argument described above. The plot of $t_{\textrm{echo}}$ versus $Q$ in the right panel of \autoref{echo_time} confirms the same.  In the bottom left panel of \autoref{Ringdown}, we present the ringdown waveforms of UCOs ($0.01<\epsilon<\epsilon_{\textrm{ph}}$). In this scenario, the gravitational wave echoes are absent. This is due to fact that with the increase of $\epsilon$, the size of the cavity  between the surface of the compact object and the maxima of the perturbation potential $V_{\textrm{max}}^{(i)}$ (located at $r_{\textrm{ph}}$ in the geometrical-optics approximation) decreases. These cavities are inefficient to effectively trap high frequency radiations ($\omega^2\approx V_{\textrm{max}}^{(i)}$). Here, the reflection coefficient of the compact object plays a crucial role. The reflection coefficient of the object is defined through the asymptotic behavior of $Z_{i}(r_*)$ as follows \cite{Maggio:2020jml}, 
\begin{equation}\label{reflection_asymtotic}
Z_{i}(r_{*})\sim e^{-i \omega r_{*}} + \mathcal{R}^{(i)} e^{i \omega r_{*}}\,,\quad{r_{*}\to \infty}~.
\end{equation}
This simply means that incoming waves from infinity are partially reflected back from the compact object's surface.  
From now on, we denote $\mathcal{R}^{(2)}$ by $\mathcal{R}$ for the sake of simplicity. We obtain the reflection coefficients $|\mathcal{R}|^2$ by solving the perturbation equations \autoref{perturb_eq} with boundary conditions \autoref{bc1} and \autoref{bc2} using Numerov algorithm \cite{ Caruso:2014kba}.\par  In the top left panel of \autoref{Reflectivity}, we present the reflection coefficient as a function of frequency for different values of $\epsilon$.  Here we can see that highly compact objects with $\eta=\eta_{\textrm{BH}}$, $\rho_{S}=\rho_{\textrm{SBH}}$ and smaller values of $Q$ acts as a perfect absorber of high-frequency waves. However, for less compact objects, this is no longer true. With the increase of $\epsilon$, the value of the reflection coefficient of high-frequency waves increases. Moreover, as the surface of the compact object coincides with the photon sphere radius $r_{\textrm{ph}}$, the object behaves like a perfect reflector regardless of its frequency (in this limit, the coefficient of $Z_1$ and $Z_2$ in \autoref{bc1} and \autoref{bc2} vanishes which results in reflecting boundary conditions $Z_1=Z_2=0$). Hence the high-frequency waves ($\omega^2\approx V_{\textrm{max}}^{(i)}$) can not be effectively trapped inside the cavity between the surface of the compact object and the maxima of the perturbation potential $V_{\textrm{max}}^{(i)}$ which explains the absence of echo in this limit.    
 In the top right panel of \autoref{Reflectivity}, we present the reflection coefficient as a function of frequency for different charge values. As evident, a neutral compact object with $\eta=\eta_{\textrm{BH}}$ and $\rho_{S}=\rho_{\textrm{SBH}}$ acts a perfect absorber of high frequency waves. However, in the presence of the charge parameter, the reflection coefficient of the compact object is non-vanishing.  \par
\begin{figure*}[t]
	\centering
	\minipage{0.48\textwidth}
	\includegraphics[width=\linewidth]{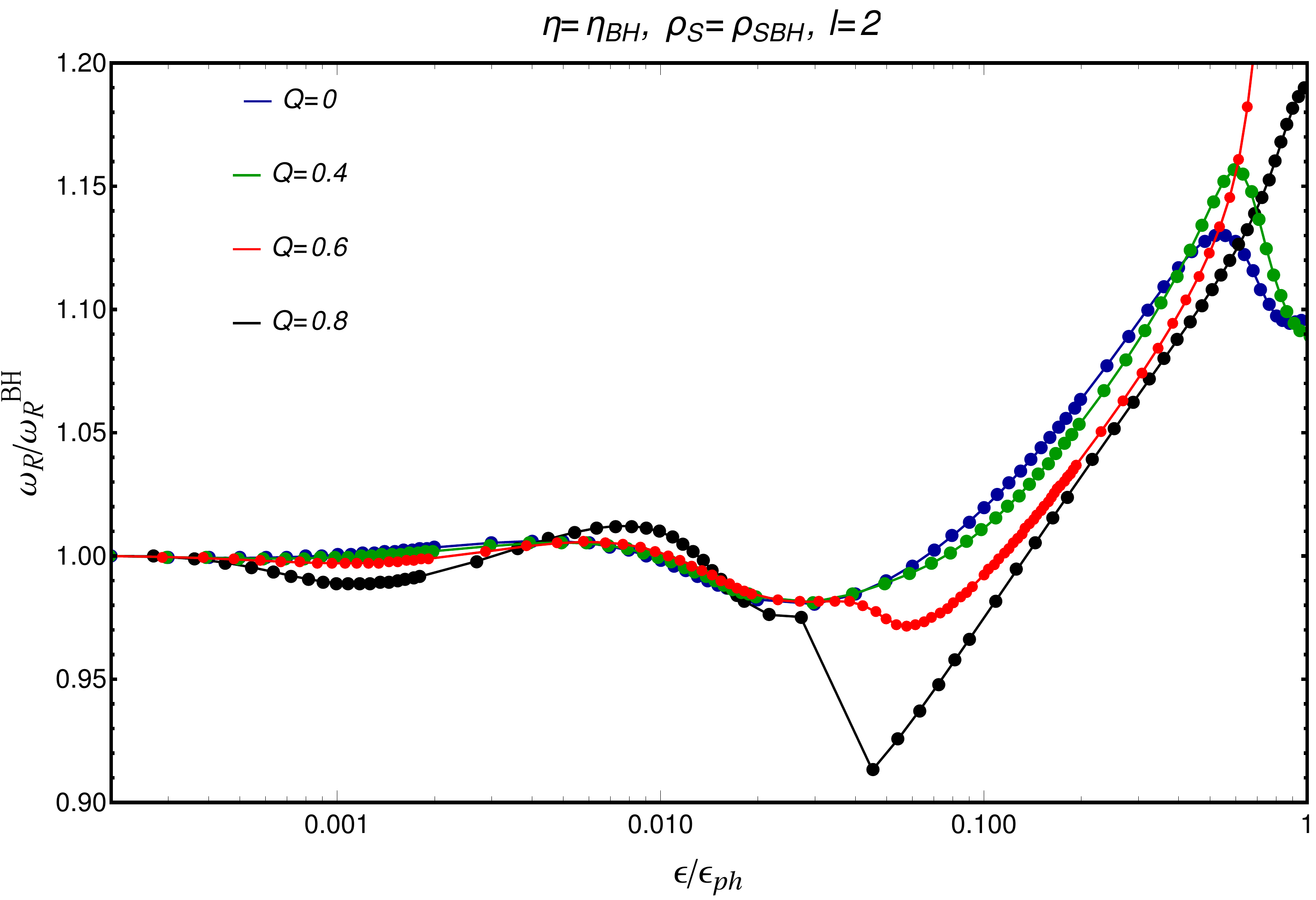}
	\endminipage\hfill
	\minipage{0.48\textwidth}
	\includegraphics[width=\linewidth]{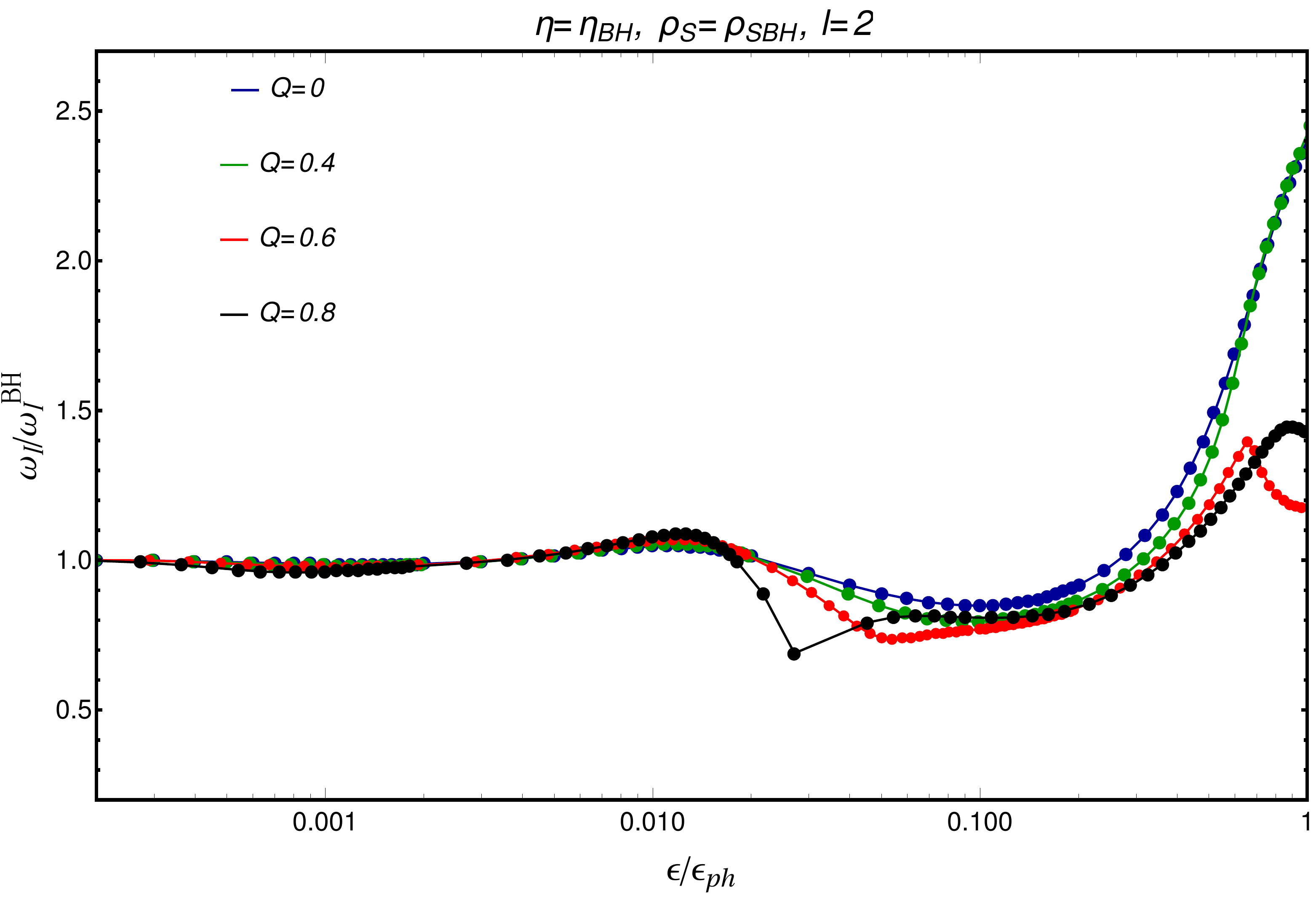}
	\endminipage
	\caption{The variation of the quantities $\omega_{\textrm{R}}/\omega_{\textrm{R}}^{\textrm{BH}}$ (left panel) and $\omega_{\textrm{I}}/\omega_{\textrm{I}}^{\textrm{BH}}$ (right panel) which represents the ratio of the fundamental quasinormal mode ($l=2$ mode) of a charged compact object with those obtained for charged black holes with respect to the compactness parameter $\epsilon$ for different values of charge $Q$ is presented. The shear viscosity and resistivity is taken as $\eta=\eta_{\textrm{BH}}=1/16\pi$ and $\rho_{S}=\rho_{\textrm{SBH}}=4\pi$. As evident, for smaller values of the compactness parameter ($\epsilon\lesssim 0.01$), the quasinormal modes of the compact object remains comparable to the black hole case. However, for less compact objects ($\epsilon\gtrsim 0.01$), quasinormal modes get modified. }\label{epsilon}
\end{figure*}
From the top panel of \autoref{Ringdown}, we find that the amplitude of the ringdown signal increases with the increase of the parameter $\eta_r\equiv1-\eta/\eta_{\textrm{BH}}$. The phenomenon can be understood by looking at the reflection coefficient's dependence on $\eta_r$ as presented in the bottom left panel of \autoref{Reflectivity}. The object behaves like a perfect absorber of high-frequency waves for smaller values of $\eta_r$ ($\eta\approx \eta_{\textrm{BH}}$). But with the increase of $\eta_r$, the value of  $|\mathcal{R}|^2$ increases. When the value of $\eta_r$ approaches unity (i.e. for $\eta\to 0$), the object becomes a perfect reflector (this is due to the fact that in the limit $\eta\to 0$, the coefficients of the $Z_1$ and $Z_2$ diverges in \autoref{bc1} and \autoref{bc2} and thus we obtain reflecting boundary conditions $Z_1=Z_2=0$). This explains the dependence of the ringdown signal on $\eta$. Moreover, from the bottom right panel of \autoref{Reflectivity}, we can see that the reflection coefficient has a very weak dependence on $\rho_r\equiv1-\rho_{S}/\rho_{\textrm{SBH}}$ for $\eta=\eta_{\textrm{BH}}$. Thus the amplitude of the signal in \autoref{Ringdown} remains practically unaltered with the change of  $\rho_r$. In order to understand this phenomena, we calculate the reflection coefficient of the high frequency waves ($M\omega\gg 1$). In this limit, we can neglect the potential term $V_{i}(r)$ in \autoref{perturb_eq}.  Similarly, we only retain the terms that are proportional to $\omega$ in \autoref{bc1} and \autoref{bc2}. Moreover, for the sake of simplicity, we consider compact objects with small charge parameter ($Q\ll M$). Under such consideration, the reflection coefficient turns out to be
	\begin{equation}
		\begin{aligned}
|\mathcal{R}|^2&=\left[\frac{1-\eta/\eta_{\textrm{BH}}}{1+\eta/\eta_{\textrm{BH}}}\right]^2-\frac{8Q}{3}\left[\frac{(1-\eta/\eta_{\textrm{BH}})}{(1+\eta/\eta_{\textrm{BH}})^2}\right]\\
&+\frac{2Q^2}{9\pi R}\left[\frac{3(\rho_{S}-\rho_{\textrm{SBH}})(1-\eta/\eta_{\textrm{BH}})+(16\pi R)\eta/\eta_{\textrm{BH}}}{(1+\eta/\eta_{\textrm{BH}})^3}\right]\\&+\mathcal{O}(Q^3)~.
		\end{aligned}
	\end{equation}
Note that, in the $Q\to 0$ limit, the expression for reflection coefficient exactly coincides with \cite{Maggio:2020jml}. Even in the presence of a small charge $Q$, the dominant contribution in the reflection coefficient comes from the first term. 
The contribution of $\rho_{S}$ comes only in the order $\mathcal{O}(Q^2)$. Moreover, note that, in the limit $\eta\to \eta_{\textrm{BH}}$, the reflection coefficient is independent of $\rho_{S}$. This explains the behaviour in \autoref{Reflectivity}.

\section{Quasinormal Normal spectra of Charged compact objects}\label{Sec_QNM}

 The quasinormal modes of the object are the eigenvalues of \autoref{perturb_eq} with the folowing boundary conditions that there are only outgoing modes ($Z_{i}\approx \exp(i\omega r_{*})$) at the asymptotic region $r_{*}\to \infty$ while it satisfies \autoref{bc1} and \autoref{bc2} at the surface of the star. 
 Here, we employ the so-called ``shooting method" to obtain the quasinormal spectrum (see \cite{Pani:2013pma,Konoplya:2011qq,Berti:2009kk} for a  detailed discussion of this method). The basic idea behind this method is to integrate the perturbation equation numerically from a point $r_0=R(1+\delta)$ (where, $ \delta<<1 $) close to the surface of the compact object to infinity, where we impose the outgoing boundary condition. 
 In order to find the behaviour of the perturbation function at $r_0$, we use the following ansatz.
 \begin{equation}\label{power_re}
 	Z_{i}(r)=(r-r_e)^{-\frac{i\omega}{2\kappa_{e}}}(r-r_c)^{\frac{i\omega}{2\kappa_{c}}}\sum_{n=0}^{N} a_{n}^{(i)}~(r-R)^n~,
 \end{equation}
 where, $\kappa_{e}$ and $\kappa_{c}$ denote the surface gravity at the position of the event, and the Cauchy horizon, respectively and $N$ is an arbitrary finite integer. 
 Using the perturbation equations \autoref{perturb_eq} along with the boundary conditions \autoref{bc1} and \autoref{bc2}, we can write the coefficients $a_{n}^{(i)}$ in terms of $a_{0}^{(i)}$. The boundary conditions dictate that the solutions at $r_0$ are a superposition of incoming and outgoing waves. Similarly, near the asymptotic infinity, we use the following power series.
 \begin{equation}\label{power_inf}
 	Z_{i}(r)=e^{k r} r^p\sum_{n=0}^{N} \dfrac{b_{n}^{(i)}}{r^{n}}~,
 \end{equation}
 where, $p=i\omega(1/\kappa_{e}-1/\kappa_{c})$. Note that, there exist two independent solutions at the infinity corresponding to $k=\pm i\omega$. This give us the behavior of $Z_{i}$ at large values of $r=r_{\textrm{inf}}$. We obtain the quasinormal modes by numerically integrating \autoref{perturb_eq} from $r_0$ to $r_{\textrm{inf}}$ and then comparing them with \autoref{power_inf} with QNM boundary condition that there are only outgoing solution at $r_{\textrm{inf}}$.\par
\begin{figure*}[t]
	\centering
	\minipage{0.48\textwidth}
	\includegraphics[width=\linewidth]{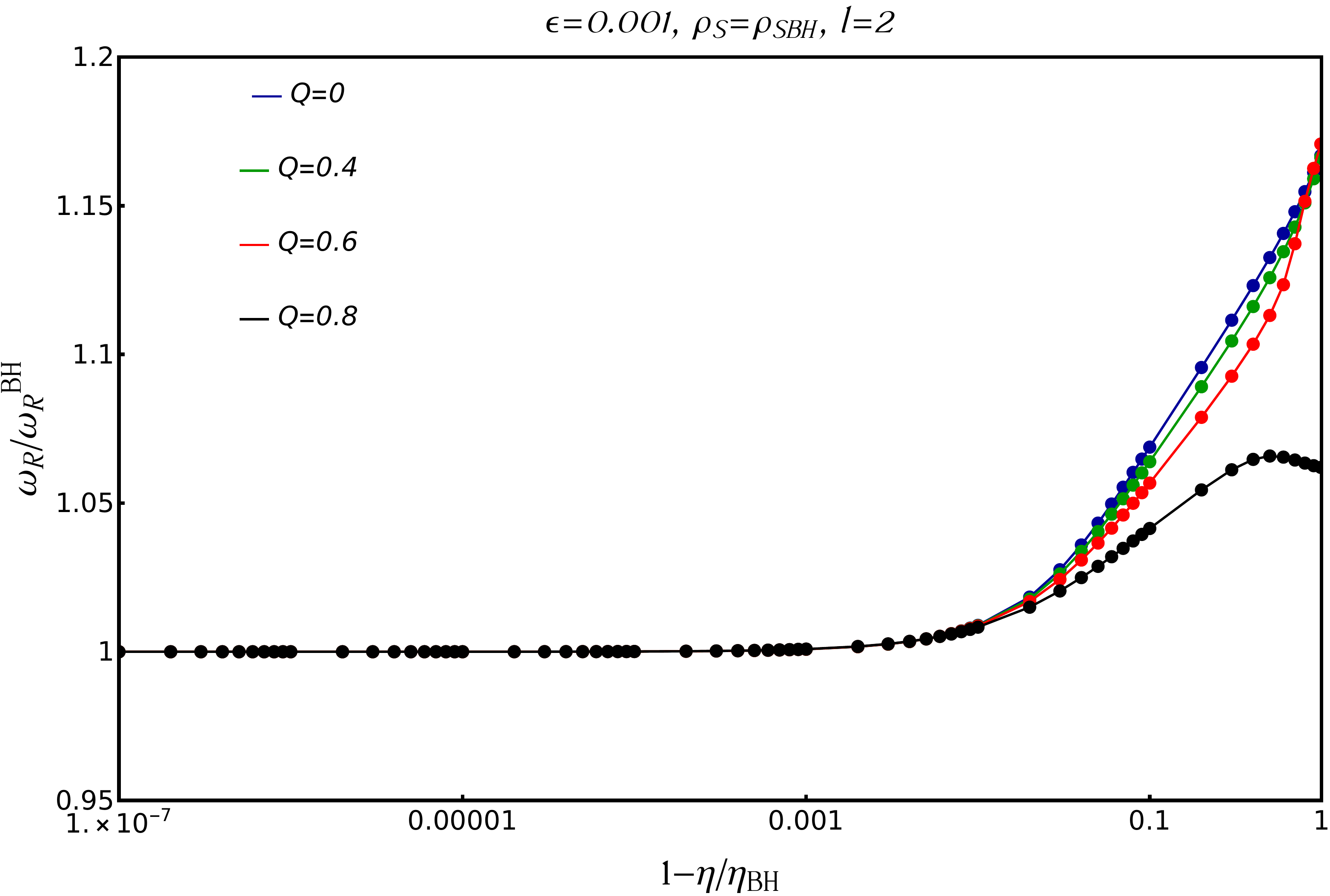}
	\endminipage\hfill
	\minipage{0.48\textwidth}
	\includegraphics[width=\linewidth]{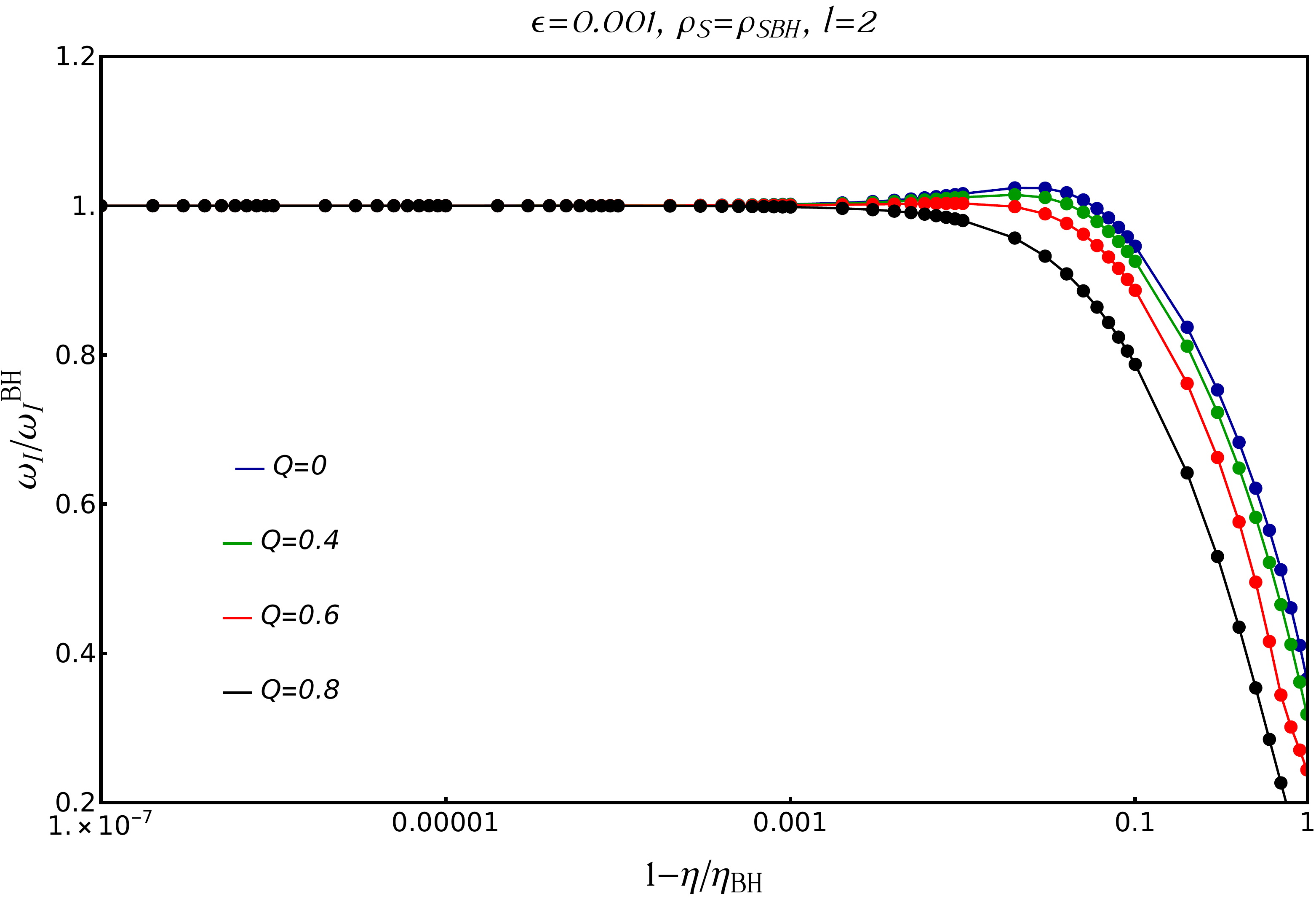}
	\endminipage
	\caption{The variation of the quantities $\omega_{\textrm{R}}/\omega_{\textrm{R}}^{\textrm{BH}}$ (left panel) and $\omega_{\textrm{I}}/\omega_{\textrm{I}}^{\textrm{BH}}$ (right panel) which represents the ratio of the fundamental quasinormal mode ($l=2$ mode) of a charged compact object with those obtained for charged black holes with respect to $\eta_r\equiv1-\eta/\eta_{\textrm{BH}}$ is plotted for different values of charge $Q$. The compactness parameter and the resistivity of the object is taken as $\epsilon=0.001$ and $\rho_{S}=\rho_{\textrm{SBH}}=4\pi$.  }\label{eta}
\end{figure*}
 In \autoref{epsilon}, we present the variation of the real (left panel) and imaginary part (right panel) of fundamental quasinormal mode ($l=2$ mode) 
as a function of the compactness parameter $\epsilon$ 
for different values of electric charge $Q$. Here, we scale the QNMs of the compact object with  $\eta=\eta_{\textrm{BH}}$ and $\rho_{S}=\rho_{\textrm{SBH}}$ relative to the value obtained in black hole scenario (the QNMs of \RN\ black holes are presented in \autoref{table3}). 
 As evident from the figure,  the QNM spectra of the ClePhOs (corresponding to $\epsilon\lesssim 0.01$)  do not differ much from that of the black holes. However, for less compact objects ($\epsilon\gtrsim 0.01$), we observe significant modification in the spectrum. 
 The phenomena can be explained through causality arguments. As discussed in the previous section, in the geometrical-optics approximation, the instability time scale $t_{\textrm{d}}$ of photon trajectories at the photon sphere governs the damping rate of the prompt ringdown signal. For \RN\ black holes, the typical value of $t_{\textrm{d}}$ is $\approx 10M$. On the other hand,  starting from the photon orbit, null geodesics completes a round trip between the photon sphere and the surface of the compact object in time $t_{\textrm{echo}}$ (see \autoref{round_trip}). Thus, the boundary will have a impact on the QNM spectrum only when $t_{\textrm{echo}}<t_{\textrm{d}}$ \cite{Maggio:2020jml}. From \autoref{echo_time}, we can see that the $t_{\textrm{echo}}$ is larger than $ 10M $ (represented by black horizontal line) for highly compact objects. Thus, the boundary does not modify the QNM spectrum for these objects.
 However, for the less compact objects, ($ \epsilon \gtrsim 0.01 $), we can easily check that $ t_{echo}<t_d $ and hence the
 boundary conditions do modify the spectrum significantly.
 \begin{figure*}[t]
 	\centering
 	\minipage{0.48\textwidth}
 	\includegraphics[width=\linewidth]{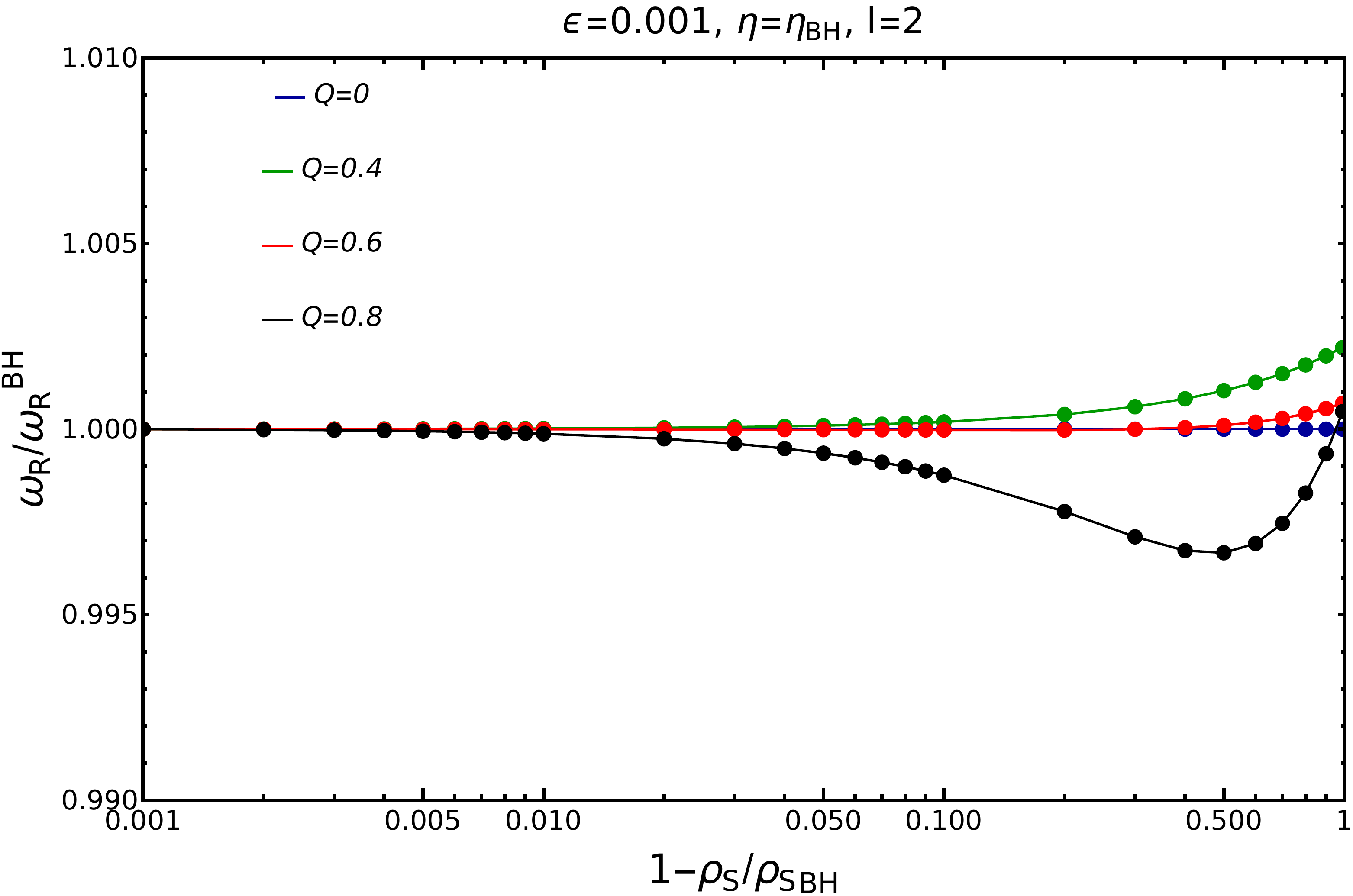}
 	\endminipage\hfill
 	\minipage{0.48\textwidth}
 	\includegraphics[width=\linewidth]{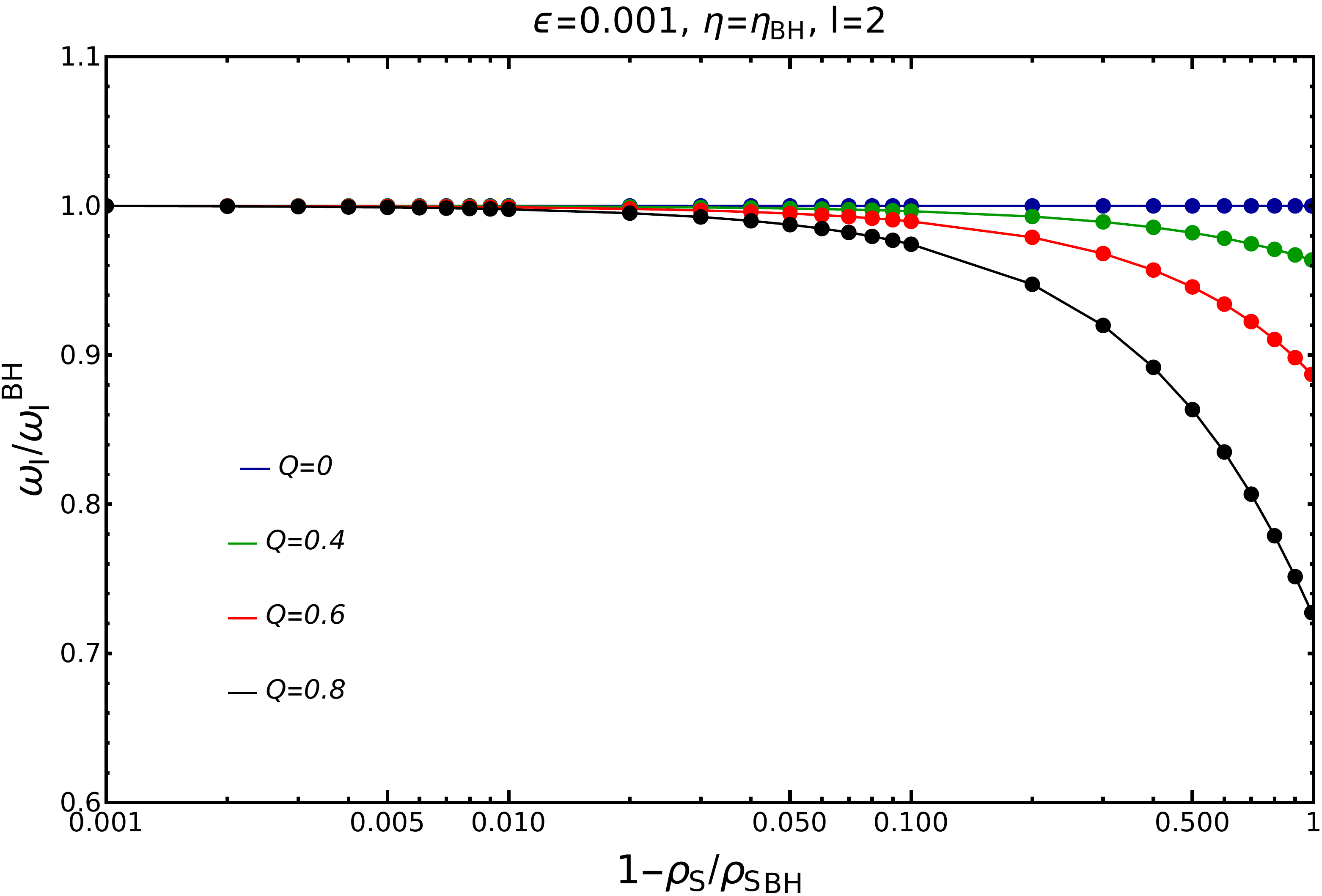}
 	\endminipage
 		\caption{The variation of the quantities $\omega_{\textrm{R}}/\omega_{\textrm{R}}^{\textrm{BH}}$ (left panel) and $\omega_{\textrm{I}}/\omega_{\textrm{I}}^{\textrm{BH}}$ (right panel) which represents the ratio of the fundamental quasinormal mode ($l=2$ mode) of a charged compact object with those obtained for charged black holes with respect to $\rho_r\equiv1-\rho_{S}/\rho_{\textrm{SBH}}$ is shown for different values of charge $Q$. The compactness parameter and the shear viscosity of the object is taken as $\epsilon=0.001$ and $\eta=\eta_{\textrm{BH}}=1/16\pi$.  }\label{rho}
 \end{figure*}
 \par
In general, the ringdown waveform of a compact object can be described as the addition of a prompt ringdown signal (primary pulse) and a series of echo pulses [48,104]. The pulses are separated by time $t_{\textrm{echo}}$ (or equivalently by a phase factor $i\omega t_{\textrm{echo}} $). If the time separations between the pulses are larger than its width (this happens for highly compact objects, see Fig.~2), there is no interference, and the total ringdown signal appears to be a sum of these individual pulses, as can be seen in the top panel of Fig.~1 \cite{Mark:2017dnq}. In this scenario, the dominant QNM is practically indistinguishable from the black hole QNM \cite{Mark:2017dnq, Bueno:2017hyj}. However, for a less compact object ($t_{\textrm{echo}}$ is small) where the separation between the echoes is smaller than the pulse width, there is no distinct echo pulse. Due to inference between the echo signals, the total ECO signal becomes a single damped sinusoid with a frequency different from that of a black hole
 \cite{Mark:2017dnq}.\par
In \autoref{eta}, we present the variation of the quantities $\omega_{\textrm{R}}/\omega_{\textrm{R}}^{\textrm{BH}}$ (left panel) and $\omega_{\textrm{I}}/\omega_{\textrm{I}}^{\textrm{BH}}$ (right panel) as a function of the parameter $\eta_r=1-\eta/\eta_{\textrm{BH}}$ for different values of charge $Q$. Here, we consider a object with compactness $\epsilon=0.0001$ and resistivity $\rho_{S}=\rho_{\textrm{SBH}}$. As evident from \autoref{eta}, when $\eta_r\lesssim 0.01$, the QNM spectum of the compact object remains practically indistinguishable from that of the black hole. However, we observe a significant modification in the spectrum when the value of $\eta_r$ approaches to unity.  
\\
The dependence of  $\omega_{\textrm{R}}/\omega_{\textrm{R}}^{\textrm{BH}}$ (left panel) and $\omega_{\textrm{I}}/\omega_{\textrm{I}}^{\textrm{BH}}$ (right panel) on the parameter $\rho_r=1-\rho_{S}/\rho_{\textrm{SBH}}$ is presented in \autoref{rho} for different values of charge $Q$.  As expected, for neutral compact objects, this parameter is irrelevant as can be seen from the figure. However, for charged objects, the QNM spectrum get modified for higher values of $\rho_r$. For real part of the QNM, the modification is less than $0.5\%$. However, the imaginary part gets significant modification in the limit $\rho_r\to 1$. 

\section{Quasinormal mode in the near extremal limit}\label{Near extremal modes}

In this section, we discuss about the quasinormal modes of the \RN\ black holes and charged compact objects in the near etremal limit $Q\to M$. Following \cite{Kim:2012mh,Chen:2012zn,Rahman:2018oso}, we use the following coordinate transformation,
	\begin{eqnarray}{\label{NEcondition}}
		& &{r}\to Q+\varepsilon \varrho\,,\quad 
		{M}\to \sqrt{Q^{2}+\varepsilon^{2}B^{2}}\,, \quad 
		{t}\to \frac{\tau}{\varepsilon}.
	\end{eqnarray}
	Here, $B$ represents the deviation from the extremity, and $\varepsilon$ is a small parameter. Under this transformation, the line element \autoref{rn_bh} can be rewritten as follows,
	\begin{equation}\label{RNNE}
		ds^{2}=-f(\varrho)~d\tau^{2}+\frac{1}{f(\varrho)}~d\varrho^{2}+Q^{2}~(d\theta^{2}+\sin^{2}{\theta}~d\phi^{2}) ~,
	\end{equation}
	where $f(\ro)=(\ro^{2}-B^{2})/{Q^{2}}$. The event horizon is located at $\ro=B$. We consider that the surface of the compact object is located at $\ro_R=B(1+\epsilon)$. By using the transformation \autoref{NEcondition} and introducing the tortoise coordinate $\ro_{*}=\log[(\ro-B)/(\ro+B)]/2\kappa$, we rewrite the perturbation equations \autoref{perturb_eq} for $Z_1$ and $Z_2$ as follows,
\begin{equation}\label{NE_pert}
\begin{aligned}
&\dfrac{d^2 Z_i(\ro_{*})}{d\ro_{*}^2}+\left[w^2-V_i(\ro_{*})\right]Z_i(\ro_{*})=0~,\\
&V_i(\ro_{*})=\frac{\sigma_i(\sigma_i+1)\kappa^2}{\sinh^2{(\kappa \ro_{*})}}
\end{aligned}
\end{equation}
where, $w=\omega/\varepsilon$, $\kappa=B/Q^2$, $\sigma_1=(l+1)$, and $\sigma_2=(l-1)$. Here, we consider terms upto $\mathcal{O}(\varepsilon^2)$. Moreover, we can write \autoref{NE_pert} as, 
\begin{equation}\label{NE_pert_1}
\begin{aligned}
&x(x-1)\frac{d^{2}Z_i(x)}{dx^{2}}+(1-\frac{3x}{2})\frac{dZ_i(x)}{dx}\\&+\left[\frac{w^2}{4\kappa^2 x}-\frac{\sigma_i(\sigma_i+1)}{4(1-x)}\right]Z_i(x)=0
\end{aligned}
\end{equation}
by introducing a new variable $x=1/\cosh^2{(\kappa \ro_{*})}$. The solution of this equation can be written in terms of hypergeometric functions as follows,
\begin{equation}\label{NE_sol}
\begin{aligned}
&Z_i(x)=(1-x)^{-\frac{\sigma_i}{2}}\bigg[A_i~x^{-\frac{i w}{2\kappa}}F_{1}(a_i,b_i,c;x)\\&+B_i~ (-x)^{\frac{i w}{2\kappa}}F_{1}(1+a_i-c,1+b_i-c,2-c;x)\bigg]
\end{aligned}
\end{equation}
where, 
\begin{equation}\label{a_b_c}
\begin{aligned}
a_i=-\frac{i w}{2\kappa}-\frac{\sigma_i}{2}+\frac{1}{2}\,,\quad{b_i}=-\frac{i w}{2\kappa}-\frac{\sigma_i}{2}\,,\quad{c}=1-\frac{i w}{\kappa}~.\nonumber
\end{aligned}
\end{equation}
Similarly, we can write the boundary conditions \autoref{bc1} and \autoref{bc2} in terms of the variable $x$. In this section, we are interested in finding the quasinormal spectrum of highly compact objects for which the surface of the compact object is located at $\ro_{*}(\ro_R)\approx\log[\epsilon]/2\kappa\ll 0$ ($x_R\approx 2\epsilon$). In this scenario, the exact expression of the boundary conditions is not needed to obtain the quasinormal modes. Near the surface of the compact object, the solution becomes
\begin{equation}\label{NE_NH}
	\begin{aligned}
		Z_i(x)&=A_i^{\textrm{in}}~e^{-i w \ro_{*}}+B_i^{\textrm{out}}~e^{i w \ro_{*}}
	\end{aligned}
\end{equation}
where, $A_i^{\textrm{in}}=A_i C_i$ and $B_i^{\textrm{out}}=B_i D_i$, and
\begin{equation}\label{NE_NH1}
	\begin{aligned}
C_i&=2^{-\frac{i w}{2\kappa}}(1+\frac{2a_i b_i}{c}\epsilon)~,\\D_i&=~ (-2)^{\frac{i w}{2\kappa}}\left(1+\frac{2(1+a_i-c) (1+b_i-c)}{2-c}\epsilon\right)~.
	\end{aligned}
\end{equation}
Here, we used the following property of the hypergeometric function $F_1(a,b,c;z)=1+abz/c$. Similarly, near the asymptotic boundary $\ro\to \infty$ ($\ro_{*} \to 0$, $x\to 1$), the solution becomes
\begin{equation}\label{NE_sol1}
	\begin{aligned}
		&Z_i(x)=(1-x)^{-\frac{\sigma_i}{2}}\bigg[\frac{A_i^{\textrm{in}}}{C_i}~x^{-\frac{i w}{2\kappa}}F_{1}(a_i,b_i,c;1)\\&+\frac{B_i^{\textrm{out}}}{D_i}~ (-x)^{\frac{i w}{2\kappa}}F_{1}(1+a_i-c,1+b_i-c,2-c;1)\bigg]~.
	\end{aligned}
\end{equation}
Demanding that there is only outgoing solution at the asymptotic boundary, the coefficient of $A_i^{\textrm{in}}$ should vanish. Making use of the following property of hypergeometric function,
\begin{equation}\label{hyper}
	\begin{aligned}
	F_{1}(a,b,c;1)=\dfrac{\Gamma(c)\Gamma(c-a-b)}{\Gamma(c-a)\Gamma(c-b)}~,
	\end{aligned}
\end{equation}
we obtain the following conditions, $c-a_i=-n$ or $c-b_i=-n$, where $n$ is an integer. Substituting the values of $a_i$, $b_i$, and $c$, we obtain the expression for quasinormal modes as 
\begin{equation}\label{NE_QNM}
		w^{i}_{n}=-i(n+\sigma_i+1)\kappa~
\end{equation}
Note that, Ref. \cite{Kim:2012mh,Chen:2012zn,Rahman:2018oso} found the same expression of quasinormal modes for near extremal \RN\ black holes \footnote{Ref. \cite{Kim:2012mh,Chen:2012zn,Rahman:2018oso} calculates the quasinormal mode of near extremal \RN\ black holes under scalar field perturbation where the perturbation equation has the same form as \autoref{NE_pert} with $\sigma_i=l$ (see \cite{Chen:2012zn}).}. Thus, even in $Q\to M$ limit,  the quasinormal spectrum of a highly compact object remains the same as the \RN\ black holes, which reassures our previous results.

\section{Detectability of GW echoes}\label{Detect}

\begin{figure}[t!]
	\centering
	\includegraphics[width=0.485\textwidth]{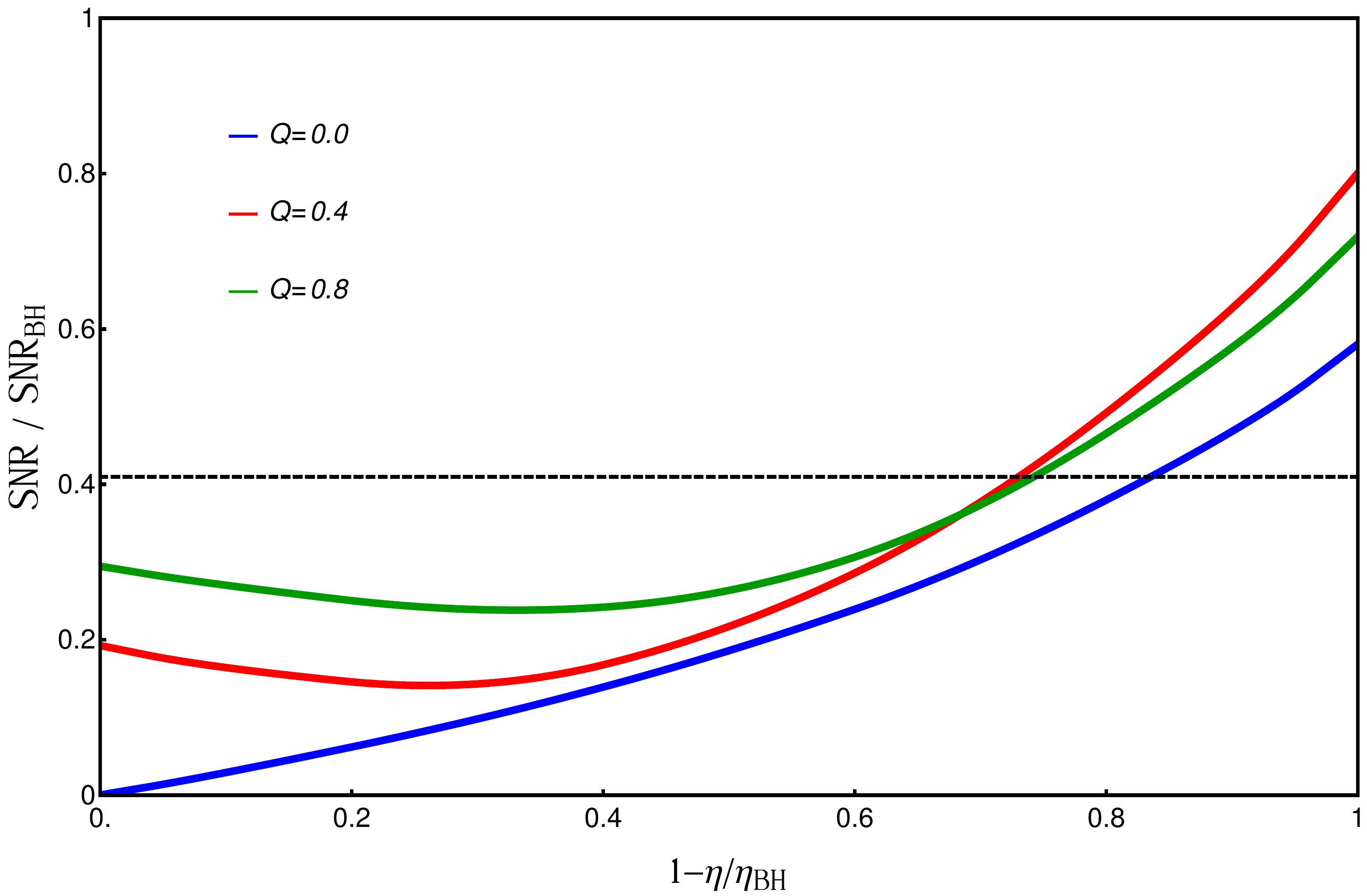}
	\caption{
		The SNR for echo signal relative to prompt ringdown is plotted as function of $\eta_r\equiv1-\eta/\eta_{\textrm{BH}}$ for different values of $Q$. Here, we take $\epsilon=10^{-8}$ and $\rho_{S}=\rho_{\textrm{SBH}}$. The black horizontal line in the plot represents the detection threshold, corresponding to SNR$=10$. We assumed that the SNR for prompt ringdown is 24.4, similar to event GW150914. 
	} 
	\label{fig:SNR}
\end{figure}

One interesting feature about the perturbation of \RN\ geometry is that Maxwell's mode  $H(r)$ appears in the quasinormal spectrum along with the gravitational modes $U(r)$ \cite{Moncrief:1975sb, PhysRevD.9.860,Chand}. As discussed in \autoref{Sec_Perturb}, the perturbation functions $H(r)$ and $U(r)$ form a system of coupled differential equations (which are related to master function $Z_1(r)$ and $Z_2(r)$ via \autoref{Z1_Z2}). Thus, excitation of one mode will inevitably excite the other. \\
The energy flux at infinity in the Maxwell and gravitational sector is proportional to $|H(r)|^2$ and $|U(r)|^2$ respectively for each multipole \cite{Cardoso:2016olt}. Hence, we calculate gravitational perturbation function $U(r)$ using \autoref{Z1_Z2}, \autoref{perturb_eq}, \autoref{bc1} and \autoref{bc2} and  discuss about the relative strength of echo signal as compared to the prompt ringdown signal of black holes. 
For this purpose, we subdivide the ringdown waveform of compact objects as $U(t)=U^{\textrm{BH}}(t)+U^{\textrm{echo}}(t)$, where $U^{\textrm{BH}}(t)$ represents the prompt ringdown signal whereas $U^{\textrm{echo}}(t)$ is the echo signal \cite{Wang:2018mlp}. Here, we adopt matched filter method to calculate the signal to noise ratio (SNR), the expression of which can be represented as follows \cite{Allen:2005fk,TheLIGOScientific:2016qqj,PhysRevD.98.044018}
\begin{equation}\label{SNR}
\textrm{SNR}=\sqrt{4\int_{0}^{\infty}\frac{|\tilde{U}(f)|^2}{S_n(f)}df}~,
\end{equation}
where, $\tilde{U}(f)=\int_{-\infty}^{\infty}\exp(2\pi i ft)U(t)dt$ is the Fourier transformation of the strain $U(t)$ and $S_n(f)$ is the one-sided power spectral density of the detector. For our study, we scale the ringdown signal to have the same amplitude as GW150914 \cite{Abbott:2016izl, LIGOdat}. We obtain $S_n(f)$ from the estimated noise curve during O1 \cite{LIGOGWOSC,LIGOASD}. The result is presented in \autoref{fig:SNR} where we plotted the relative SNR of the echo signal to prompt ringdown as a function of $\eta_r$ for different values of $Q$. Here, we choose $\eta=10^{-8}$ and $\rho_{S}=\rho_{\textrm{SBH}}$. In the plot, the black horizontal line represents the threshold of detection, corresponding to SNR$=10$. Moreover, we have assumed that the SNR for the prompt ringdown is $24.4$ similar to event GW150914 \cite{TheLIGOScientific:2016qqj,SNRdat}. As can be seen from the plot, for an uncharged configuration, LIGO detectors will discover GW echo for $\eta_r\gtrsim 0.8$ ($\eta\lesssim 0.2 \eta_{\textrm{BH}}$). However, the parameter range for which detection is possible, gets widened in presence of charge parameter ($\eta_r\gtrsim 0.7$ for $Q=0.4$ and $Q=0.8$). However, the SNR has a very weak dependence on the compactness parameter and resistivity.  

\section{Conclusion}\label{Conclusion}

 The rapid developments of GW detectors made it possible to detect more and more events with a large signal-to-noise ratio, and we are hopeful that GW spectroscopy will eventually enable us to probe physics in most extreme conditions,  from the strong gravity regime \cite{Berti:2015itd,TheLIGOScientific:2016src,Yunes:2016jcc} to the very early universe \cite{Bethke:2014oos,Boyle:2006zz}. It also pave the way to test different predictions of GR (e.g. the no-hair theorem)\cite{Cardoso:2016ryw} and to find quantum mechanical effects near the horizon \cite{Cardoso:2016rao,Barausse:2014tra,Dey:2020wzm,Basak:2020jzr}. Recently, it has been noted that GW spectroscopy can also help us detect the existence of horizon as the ringdown properties of black holes generally differs from horizonless compact objects \cite{Cardoso:2017cqb,Cardoso:2016oxy,Cardoso:2016rao,Cardoso:2019rvt,Maggio:2020jml, Mark:2017dnq, Micchi:2019yze,Micchi:2020gqy}. This can be seen from \autoref{Ringdown} where we have compared the ringdown properties of charged non-rotating compact objects and black holes. We obtain a repetitive signal for (at least!) extremely compact objects at later times, which is in contrast to the black hole scenario where there are decaying power-law tails \cite{PhysRevD.34.384,PhysRevD.49.883,PhysRevD.49.890,PhysRevLett.84.10}. The perturbation theory very accurately describes this ringdown phase. However, first, we have to set up an appropriate boundary to study the response of these objects to external perturbations, which is extremely tricky since the internal structure of these objects are largely unknown \cite{Mark:2017dnq, Micchi:2019yze,Micchi:2020gqy}. Following \cite{Maggio:2020jml}, we have employed a membrane paradigm to obtain the quasinormal spectrum of these objects in a model-independent way. The paradigm tells us that the surface of the compact object essentially behaves like a viscous fluid membrane to an outside observer \cite{1982mgm..conf..587D, Thorne, MacDonald:1982zz, PhysRevD.33.915}. We have shown that the perturbations of these objects in the odd party sector depend on the compactness $\epsilon$, the shear viscosity $\eta$ and resistivity $\rho_{\textrm{S}}$ of the membrane along with its mass and charge. \\
Our study shows that both ringdown waveform and quasinormal modes carry information about the nature of the object. In particular, we have shown that gravitational wave echoes are present in the case of ClePhOs (objects with compactness parameter $\epsilon\lesssim 0.01$).  However, the echo time $t_{\textrm{echo}}$ decreases with the increase of $\epsilon$. Finally, for UCOs (objects with  $0.01\lesssim\epsilon\lesssim \epsilon_{\textrm{ph}}$), either the echos are absent, or the waveform is a superposition of prompt ringdown and the first echo. The echo time increases with the increase of charge parameter $Q$. However, the effect is most significant in the limit $Q\approx M$. The shear viscosity $\eta$ of the membrane controls the modulation of the echos. The amplitude of the ringdown signal decreases with the increase of $\eta$. Lastly, the resistivity $\rho_{S}$ has a small effect on the ringdown signal. The phenomena can be explained by studying the optical properties of the charged membrane, where the electromagnetic perturbation can be treated as the incidence of an electromagnetic field on a charged sphere. When the electromagnetic field passes through a medium, it interacts with the internal charge and current density. Hence, it give rise to complex frequency-dependent optical constants, like complex refractive index $N=n+i k$, skin depth $\delta$, attenuation coefficient $\alpha$, and complex conductivity function $\sigma$ \cite{Kittel2004}. Here, $n$ and $k$ represent the ordinary refractive index and extinction coefficient, respectively. Both of these parameters depend on the complex conductivity function. Moreover, for normal incidence of the electromagnetic wave, the reflection coefficient becomes $|\mathcal{R}|^2=((n-1)^2+k^2)/((n+1)^2+k^2)$ which also depends on the conductivity. In our opinion, an analogical situation can be found for the membrane paradigm too. However, solving Maxwell's equation in our case is quite cumbersome, and for the moment, we left it for future work to make a more concrete connection.\ \par
For larger values of the compactness parameter, the QNMs could be a distinguisher between a black hole and a compact object. This can be seen from \autoref{epsilon}. For small values of compactness ($\epsilon\lesssim 0.01$) and charge parameter ($Q$), the quasinormal spectrum of these objects is practically indistinguishable from those charged black holes. The interior of the object will have a significant effect on the QNM spectrum if $t_{\textrm{echo}}$  is less than the decay time scale $t_{\textrm{d}}$ of a photon at the photon orbit \cite{Maggio:2020jml} which is not the case in the above-mentioned scenario. However, with the increase of compactness parameter, $t_{\textrm{echo}}$ becomes less than $t_{\textrm{d}}$ and we notice a modification in the QNM spectrum.  Similarly, we show that the QNM spectrum gets modified as the value of the shear viscosity $\eta$ and resistivity $\rho$ shifts away from $\eta_{\textrm{BH}}$ and $\rho_{\textrm{SBH}}$ respectively.  However, in the limit $Q\to M$, the quasinormal modes of highly compact objects become independent of the shear viscosity and resistivity. The quasinormal modes become purely imaginary in this scenario, similar to the near-extremal \RN\ black holes \cite{Kim:2012mh,Chen:2012zn,Rahman:2018oso}.\par
 Moreover, we have shown that both the compactness parameter and charge play a significant role in determining the reflection coefficient of the object. A object with $\eta=\eta_{\textrm{BH}}$ and $\rho_{S}=\rho_{\textrm{SBH}}$ behaves like a perfect absorber of high frequency waves for smaller values of $\epsilon$ and $Q$. However, with the increase of these two parameters, the reflection coefficient of the object is non-vanishing. In particular, when the surface of the compact object coincides with the photon sphere radius, it behaves like a perfect reflector. This behaviour is also true in the limit $\eta\to 0$.  \par
We discuss the detectability of the echo signal in \autoref{Detect}. Our study reveals that the  LIGO detectors will detect the echo signal from an uncharged compact object for the parameter range $\eta\in(0.8, \eta_{\textrm{BH}})$. Interestingly, the parameter range for which the detection is possible increases for a charged compact object. The SNR has a  weak dependence on the compactness parameter and the resistivity. \par
It is straight forward to extend the calculation in even parity sector. However, in this scenario, the non-vanishing component of $\vec{E}^\parallel_{\textrm{FIDO}}$ is $\vec{E}^\theta_{\textrm{FIDO}}$ (see \autoref{E_phi} and \autoref{resistivity_1}). Moreover, there will be a surface current along $\theta$ direction, $j^\theta$ as the membrane has a velocity component $\delta u^\theta$ along this direction (see the Appendix of \cite{Maggio:2020jml}). Furthermore, as discussed in \cite{Maggio:2020jml}, the ringdown properties will depend on an additional parameter namely on its the bulk viscosity $\zeta$ in this scenario. \par
Since our analysis is quite generic, it can describe the ringdown properties of a large variety exotic compact objects \cite{Mathur:2005zp,Bena:2007kg,Bena:2013dka,Skenderis:2008qn,Mathur:2008nj,Chapline:2000en,Mazur:2004fk,Visser:2003ge,Brito:2015yga, Liebling:2012fv, Seidel:1993zk,Gimon:2007ur,Buoninfante:2019swn,Damour:2007ap}. A possible extension of our study is to analyze the ringdown properties of magnetic black holes \cite{Maldacena:2020skw} and compact objects like Magnetars \cite{Mereghetti:2008je}. It is well known that electrically charged astrophysical objects get neutralized rather quickly by accreting ionized plasma and other quantum mechanical processes like Schwinger's pair production and Hawking radiation \cite{Cardoso:2016olt, Ghosh:2020tdu}. However, magnetic black holes are relatively long-lived (see \cite{Ghosh:2020tdu} for further discussion). Moreover, in certain parameter spaces, the magnetic field near the horizon could be quite large so that it can restore electroweak symmetry. Another possible extension is to consider the effect of spin. However, these are beyond the scope of this paper.   
\section*{Acknowledgements}
We thank Kabir Chakravarti for useful discussion. M.R is supported by the postdoctoral fellowship (MIS/IITGN/PD-SCH/201415-006) by IIT-Gandhinagar. A.B. is supported by  Start Up Research Grant (SRG/2020/001380) by Department of Science \& Technology Science and Engineering Research Board (India).
\appendix
\section{Boundary condition at the surface of the object}\label{App_RN_perturb}

We assume that at each point of the timelike stretched horizon $\mathcal{S}$; there is a fiducial observer (FIDO) moving with four-velocity $U^\mu$. The observer is at rest with respect to the compact object and making the measurements in his local Cartesian coordinate system. 
\begin{equation}\label{local_coordinate}
	\begin{aligned}
	&\hat{e}_t=\partial_\tau=\alpha^{-1}\partial_t\,,\qquad\hat{e}_r=\sqrt{f}\partial_r\\
&\hat{e}_\theta=\dfrac{1}{r}\partial_\theta\,,\qquad\hat{e}_\phi=\dfrac{1}{r\sin\theta}\partial_\phi~,
	\end{aligned}
\end{equation}
where, $\alpha=\sqrt{f}$ is called the lapse function. The normal to the strechted horizon  $n^\mu$ is a spacelike i.e. $n_\mu n^\mu=1$. For the spacetime given by \autoref{rn_bh}, the normal to $\mathcal{S}$ and the four velocities of the FIDO can be represented as follows, 
\begin{eqnarray}\label{normal}
	&n^\mu=(0,\sqrt{f},0,0)\,,\qquad U^\mu=(\dfrac{1}{\sqrt{f}},0,0,0)~.
\end{eqnarray}
Using \autoref{normal}, we can obtain the expression for extrinsic curvature from the relation $K^a_{~b}=h^c_{~b}n_{a;c}$ which is given as follows, 
\begin{equation}\label{extrinsic_curveture}
	\begin{aligned}
	&K_{tt}=-\dfrac{1}{2}\sqrt{f(r)}f'(r)\\& K_{\theta\theta}=\dfrac{K_{\phi\phi}}{\sin^2\theta}=r\sqrt{f(r)}~.
	\end{aligned}
\end{equation}
Replacing these expressions in \autoref{Membrane_paradigm} and \autoref{ST_tensor}, we obtained the expression for density and pressure as given in \autoref{rho_p}.
\subsection{First Boundary condition at the surface of the object}\label{App_boundary_condi}

Upon perturbation, the only non-vanishing components of $\delta g_{\mu\nu}$ in the axial sector are $\delta g_{t\phi}$ and $\delta g_{r\phi}$ as given by \autoref{perturb_tensors}. As discussed in \autoref{Sec_Perturb}, upto the first order of perturbations, the components of $n^\mu$ remains unchanged. Thus the non vanishing components  of the $\delta K_{\mu\nu}$ becomes
\begin{equation}\label{extrinsic_curveture_pert}
	\begin{aligned}
			\delta K_{t\phi}&=\frac{1}{2}\sin \theta e^{-i  \omega t} \sqrt{f}\left(h_0'+i \omega  h_1\right) \partial_\theta P_l(\cos\theta )\\ 
		\delta K_{\theta\phi}&=\frac{1}{2} h_1(r) \sin (\theta ) e^{-i t \omega } \sqrt{f}\bigg(2 \cot (\theta ) \partial_\theta P_l(\cos\theta)\\& +l (l+1) P_l(\cos\theta )\bigg)~.
	\end{aligned}
\end{equation}
\\
 whereas the 4-velocity of the FIDO takes the following form
\begin{equation}\label{pert_velocity}
U^\mu=(\dfrac{1}{\sqrt{f}},0,0,\delta u^\phi)~.
\end{equation}
Substituting \autoref{extrinsic_curveture_pert} and \autoref{perturb_tensors} in \autoref{Membrane_paradigm}, we obtain the expression for nonvanishing component of the energy momentum tensor of the membrane as follows,
\begin{widetext}
\begin{equation}\label{st_tensor_pert_1}
	\begin{aligned}
		&\delta \tau_{t\phi}=\frac{\sin (\theta ) e^{-i  \omega t} \partial_\theta P_l(\cos\theta ) \left(r h_0(r) f'(r)+f(r) \left(-i r \omega  h_1(r)-r h_0'(r)+4 h_0(r)\right)\right)}{16 \pi  r \sqrt{f(r)} }\\ 
		&\delta \tau_{\theta\phi}=-\frac{h_1(r) \sin (\theta ) e^{-i t \omega } \left(2 \cot (\theta )\partial_\theta P_l(\cos\theta )+l (l+1) P_l(\cos\theta )\right)}{16 \pi  \sqrt{f(r)}}~.
	\end{aligned}
\end{equation}
\end{widetext}
Again by substituting \autoref{pert_velocity} and \autoref{perturb_tensors} in \autoref{ST_tensor}, we obtain another expression for the energy-momentum tensor of the membrane in terms of dissipative fluid parameters as follows,
\begin{equation}\label{st_tensor_pert_2}
	\begin{aligned}
		\delta \tau_{t\phi}&=\sin (\theta ) \bigg[-R^2 \text{$\delta $u}^{\phi } \sqrt{f(R)} \sin (\theta ) (P+\rho )\\&-\rho  h_0(r) e^{-i t \omega } \partial_\theta P_l(\cos\theta )\bigg]\\ 
		\delta \tau_{\theta\phi}&=\eta  \left(-R^2\right) \sin ^2\theta ~ \partial_\theta\text{$\delta $u}^{\phi }~.
	\end{aligned}
\end{equation}
By equating $t\phi$ component of \autoref{st_tensor_pert_1} and \autoref{st_tensor_pert_2}, we obtain the expression for $\delta u^\phi$ as 
\begin{equation}\label{u_phi}
	\begin{aligned}
	\delta u^\phi=\frac{ e^{-i  \omega t } \partial_\theta P_l(\cos\theta ) \left(-h_0 f'+f \left(h_0'+i \omega  h_1\right)\right)}{r \sin (\theta )\sqrt{f} \left(r f'-2 f\right)}
	\end{aligned}
\end{equation}
Replacing \autoref{u_phi} in $\theta\phi$ component of \autoref{st_tensor_pert_2} and equating it with the $\theta\phi$ component of \autoref{st_tensor_pert_1}, we obtain the following expression,
\begin{equation}\label{bc_prepre}
	\begin{aligned}
\frac{-h_0(r) f'(r)+f(r) \left(h_0'(r)+i \omega  h_1(r)\right)}{f(r) \left(r f'(r)-2 f(r)\right)}=-\frac{h_1(r)}{16 \pi  \eta  r }
	\end{aligned}
\end{equation}
By substituting $h_0=(if(r)/\omega)\partial_r(h_1 f(r))$ and redefining the function $h_1(r)=(2 i \omega r U(r))/\mu f(r)$ and $\delta F_{\theta\phi}=l(l+1) H(r) $, we obtain the boundary condition for $U(r_*)$
\begin{widetext}
	\begin{equation}
		U'\left(r_*\right)\bigg{|}_{r_*(R)}=\bigg[\frac{\mu  Q f(r) H\left(r\right)}{P_h(r)}-\left(\frac{f(r) \left(r \left(\left(\mu ^2+2\right) r-6 M\right)+4 Q^2\right)}{2 r P_h(r)}+\frac{i \omega }{16 \pi  \eta }\right)U\left(r\right)\bigg]_{r=R}
	\end{equation}
	Note that, this equation coincides with \autoref{grav_pert}.
\end{widetext}
\bibliography{Reference_1}

\bibliographystyle{utphys1}
\end{document}